\documentclass[12pt]{article}
\usepackage{latexsym}
\newcommand{\bba}{\begin{eqnarray}}
\newcommand{\eea}{\end{eqnarray}}
\newcommand{\bb}{\begin{equation}}
\newcommand{\ee}{\end{equation}}
\newcommand{\bban}{\begin{eqnarray*}}
\newcommand{\eean}{\end{eqnarray*}}

\newcommand{\dddotV}{\buildrel{...}\over V}
\newcommand{\ddddotV}{\buildrel{....}\over V}
\begin{document}

\begin{titlepage}

\title{Stress-energy tensor in colliding plane wave space-times: An
approximation procedure}

\author{\null}
\author{  {\sc Miquel  Dorca}\footnote{
E-mail: mdorca@rainbow.uchicago.edu}$\;\;$\footnote{
Present adress: Box 1843, Department of Physics, 
               Brown University, Providence RI 02912
                                              } \\ 
          {\small\em  Enrico Fermi Institute}   \\
          {\small\em  The University of Chicago}  \\
          {\small\em  5640 Ellis Avenue}        \\
          {\small\em  Chicago IL 60637}    }

\date{October 31, 1997}

\maketitle

\begin{abstract}

In a recent work on the quantization of a massless scalar field in a
particular colliding plane wave space-time, 
we computed the vacuum expectation value of the
stress-energy tensor 
on the physical state which corresponds to the Minkowski vacuum before
the collision of the waves. We did such a calculation
in a region close
to  both the Killing-Cauchy horizon and the {\em folding singularities}
that such a
space-time contains.
In the present paper, we give a suitable
approximation procedure to compute this  expectation value, in the
conformal coupling case, throughout
the causal past of the center of the collision. This will allow us
to approximately study the evolution of such an  expectation value  from the
beginning of the collision until the
formation of the Killing-Cauchy horizon.
We start with a null expectation value before the arrival of the
waves, which then acquires nonzero values at the beginning of the collision
and grows unbounded towards the Killing-Cauchy horizon. The
value near the horizon is compatible with our
previous result, which means that such an approximation 
may be applied to other colliding plane wave space-times. Even with this 
approximation,
the initial modes propagated into the interaction region contain a
function which cannot be calculated exactly and to ensure the correct
regularization of the stress-energy tensor with the point-splitting
technique, this function must be given up to adiabatic order four of 
approximation.

\end{abstract}



\end{titlepage}

\section{Introduction}

Although exact {\it gravitational plane waves}
are very simple time dependent plane symmetric solutions to Einstein's
equations \cite{bel26}, they show some nontrivial global features
which are essentially due to the non-linearity of General
Relativity. One of these features is the absence of a global Cauchy surface,
which is a consequence of the focusing effect that the waves exert on
null rays. Another feature, related to the absence of a global Cauchy
surface, is the presence of a Killing-Cauchy horizon which may be physically
understood as the caustic produced by the focusing of null rays \cite{pen65}. 
The inverse of
the focusing time is a measure of the strength of the wave, thus for an 
Einstein-Maxwell plane wave such inverse
time equals the electromagnetic energy per unit surface of the wave. This 
makes exact plane waves very
different from their linearized counterparts, which have no focusing points 
and admit a globally hyperbolic
space-time structure. One expects that exact plane waves may be relevant for 
the study of the strong time
dependent gravitational fields that may be produced in the collision of black 
holes \cite{dea79,fer80} or to
represent travelling waves on strongly gravitating cosmic strings 
\cite{gar89-90}. In recent years these
waves have been used in classical general relativity to test some conjectures 
on the stability of Cauchy
horizons
\cite{ori92,yur93}, and in string theory to test classical and quantum string 
behaviour in strong
gravitational fields \cite{veg84-90,veg91,jof94}. Their interest also stems 
from the fact that plane waves are a
subclass of exact classical solutions to string theory 
\cite{ama84-88,hor90,tse93fes94rus95}. 
 
When plane waves
are coupled to quantum fields the effects are rather trivial since they 
produce neither vacuum polarization
nor the spontaneous creation of particles. In that sense these waves behave 
very much as electromagnetic or
Yang-Mills plane waves in flat space-time \cite{des75,gib75}. Still the 
classical focusing of geodesics has a
quantum counterpart: when quantum particles are present the quantum field 
stress-energy tensor between
scattering states is unbounded at the Cauchy horizon, i.e. where classical 
test particles focus after colliding
with the plane wave
\cite{gar91}. This suggests that the Cauchy horizon of plane waves may be 
unstable under the presence of quantum
particles. The classical instability of the null Cauchy horizons of plane 
waves is manifest when non-linear
plane symmetric gravitational radiation collides with the background wave, 
i.e. when {\it two plane
waves collide}. In this case the focusing effect of each wave distorts the 
causal structure of the space-time
near the previous null horizons and either a spacelike curvature singularity 
or a new regular Killing-Cauchy
horizon is formed. However, it is generally believed that the Killing-Cauchy 
horizons of the colliding plane
wave space-times are unstable in the sense that ``generic" perturbations will 
transform them into spacelike
curvature singularities. In fact, this has been proved under general plane 
symmetric perturbations
\cite{yur88-89}. Also exact colliding plane wave solutions with classical 
fields are known that have
spacelike curvature singularities and which reduce, in the vacuum limit, to 
colliding plane wave solutions
with a regular Killing-Cauchy horizon
\cite{cha87}.
 
{\it Colliding plane wave space-times} are some of the simplest dynamical 
space-times and, as such, they have been
used as a testing ground for some problems in classical 
general relativity such as 
the just mentioned stability of
the Killing-Cauchy horizons, or the cosmic censorship hypothesis \cite{yur93}. 
Note that in a colliding plane
wave space-time with a Killing-Cauchy horizon inequivalent extensions can be 
made through the horizon, and this
implies a breakdown of predictability since the geometry beyond the horizon is 
not uniquely determined by the
initial data posed by the incoming colliding plane waves. The singularities in 
colliding wave space-times are
also different from the more familiar cosmological and black hole 
singularities which originate from the
collapse of matter since they result from the non-linear effects of pure 
gravity. The type of singularities
also differs in the sense that these are all-encompassing, i.e. all timelike 
and null geodesics will hit the
singularity in the future.
 
In three previous papers  \cite{dor93,dor94,dor96} we studied the
interaction of massless scalar quantum fields with a gravitational background 
which represents the head on
collision of two linearly polarized shock waves followed by trailling 
gravitational radiation which focus into
a Killing-Cauchy horizon. The space-time is divided 
into four distinct regions: a
flat space region (which represents the initial flat region before the waves 
collide), two single plane wave
regions (the plane waves before the collision) and the interaction region 
which is
bounded by the previous three regions and a regular Killing-Cauchy horizon. 
Each single plane wave region contains a type of topological singularity
usually referred as {\em folding singularity} which is a remnand of the
coordinate singularity that a free gravitational plane wave develops as a
consequence of its focusing properties over null geodesics \cite{gri91}.
The interaction region is locally
isometric to a region inside the event horizon of a Schwarzschild black hole 
with the Killing-Cauchy horizon
corresponding to the event horizon \cite{fer87cha86,hay89}.
The presence of the Killing horizon made possible the definition of a natural 
preferred ``out" vacuum state
\cite{kay91} and it was found in \cite{dor93,dor94} that the initial flat
vacuum state contains a  
spectrum of ``out" particles. In the
long wavelength limit the spectrum is consistent with a thermal spectrum at a 
temperature which is inversely
proportional to the focusing time of the plane waves. Of course, the 
definition of such ``out" vacuum is not
possible when we have a curvature singularity (i.e. in the ``generic" case) 
instead of an horizon, whereas a
physically meaningful ``in" vacuum may be defined in all colliding plane wave 
space-times.
 
In reference \cite{dor96} we computed the expectation value of the
stress-energy tensor of  
the quantum field in a region
near both the horizon  and the folding singularities in the initial
flat space vacuum. We found, not
surprisingly, that the stress-energy tensor is unbounded at the horizon. The 
specific form of this divergence
suggests that when the backreaction is taken into account the horizon will 
become a space-time singularity,
i.e. the Killing-Cauchy horizon is unstable under vacuum polarization. Note 
that this is a non perturbative
effect but the result of the nonlinearity of gravity, since gravitational 
waves in the linear approximation
do not polarize the vacuum. In fact the vacuum stress-energy tensor of a 
quantum field in a weakly inhomogeneous
background was computed by Horowitz \cite{hor80}, and it is easy to see that 
such a tensor can be written in
terms of the linearized Einstein tensor only \cite{cam94}, which vanishes for 
gravitational waves.
 
The non perturbative evaluation of the expectation value of the stress-energy 
tensor of a quantum field in a
dynamically evolving space-time is generally a difficult task.
Even when the exact modes of the quantum field equation are known it may not 
be possible to perform the mode
sums in order to get the quantum field two-point function or, more precisely, 
the Hadamard function, which is
the key ingredient in the evaluation of the stress-energy tensor. In our 
colliding wave space-time we do not
even know the exact solution of the modes in the interaction region (a similar 
situation is produced in the
Schwarzschild case \cite{can80sci81,fro85tho86}). In the previous
paper \cite{dor96} we managed to calculate these mode sums using the
fortunate fact that the geometry of the colliding space-time is such
that the initial modes which come from the flat region are strongly
blueshifted in their frequency in the interaction region near both the
horizon and the folding singularities. 
In the present paper we will deal with the problem of calculating
these mode sums for a more generic colliding plane wave space-time.
Nevertheless, for simplicity, we will consider the particular
space-time described above and we will discuss the possible
generalizations to other colliding plane wave space-times. The main goal
of the paper will be the introduction of a suitable approximation to 
calculate such mode sums in the causal past of the center of the
collision, which will allow us to approximately compute the evolution  of the
stress-energy tensor from the beginning of the collision until the
formation of the Killing-Cauchy horizon. In fact, in the region close
to the horizon we will
recover the behaviour of the stress-energy
tensor computed in \cite{dor96}. We will see that, in order to
propagate the initial flat mode throughout the causal past of the
center of the collision, only a subset of the Cauchy data is required
and furthermore the behaviour of the differential equations satisfied
by these modes may be conveniently approximated. We
will see that these approximations can be also recovered by a
convenient change in the space-time geometry throughout the causal
past of the center of the collision. In fact, in order to evaluate
the mode sums involved in the Hadamard function and to correctly
substract the nonphysical divergences, with the point-splitting
prescriptions, it is necessary to work with approximations in the
space-time geometry rather than in the differential equations.
 
The plan of the paper is the following. In section 2 the geometry of the 
colliding plane wave space-time is
briefly reviewed. In section 3 the mode
solutions of the scalar field equation are given for the four different 
regions of the space-time, it is only
in the interaction region that exact solutions for these modes cannot be 
found and therefore we develop a suitable approximation in order to
find them, first by doing approximations in the differential equations
and then by conveniently changing the space-time geometry.
In section 4
the point splitting technique is reviewed and adapted to 
the computational purposes of 
this paper.
In section 5 this technique is used to regularize the Hadamard 
function and
calculate its value by a mode sum in the causal past of the center
of the collision. Finally, in 
section 6 the expectation
value of the stress-energy tensor near the horizon is calculated. A summary 
and some consequences of
our results, such as the backreaction problem, the quantum instability of the 
Killing-Cauchy horizon and the
generality of these results are discussed in section 7. In order to keep the 
main body of the paper reasonably
clear, some  of the technical details of the calculations have been
left to the Appendices.

\section{Description of the geometry}

We will consider in this paper an example of a colliding plane
wave space-time where instead of a curvature singularity a
Killing-Cauchy horizon is produced. 
It represents a collision of two pure gravitational shock waves
followed by trailing radiation, with an interaction region which is
locally isometric to a region of the interior of a Schwarzschild
black hole, where the Killing-Cauchy horizon corresponds to the black
hole event horizon \cite{fer87cha86,hay89}. The space-time
contains four space-time regions (see Fig. 1), given by

\bb \left.ds^2_{IV}\right.=4L_1L_2dudv-dx^2-dy^2,  \label{eq:ibIV}\ee
\bb \left.ds^2_{III}\right.=4L_1L_2(1+\sin v)^2dudv-
{1-\sin v\over 1+\sin v}\, dx^2-(1+\sin v)^2\cos ^2v\, dy^2,\label{eq:ibIII}\ee
\bb \left.ds^2_{II}\right.=4L_1L_2(1+\sin u)^2dudv-
{1-\sin u\over 1+\sin u}\, dx^2-(1+\sin u)^2\cos ^2u\, dy^2,\label{eq:ibII}\ee
\bb \left.ds^2_{I}\right.=4L_1L_2[1+\sin (u+v)]^2dudv-
{1-\sin (u+v)\over 1+\sin (u+v)}\, dx^2-[1+\sin (u+v)]^2\cos ^2(u-v)\,
dy^2,\label{eq:ibI}\ee
where for convenience we have used $u$ and $v$ as dimensionless
null coordinates, and where  $L_1$ and $L_2$, are length
parameters such that $L_1L_2$ is directly related to the focusing time
of the collision, i.e. to the inverse of the strength of the waves,
which is a measure of the amount of nonlinearity of the gravitational
waves \cite{gri91}.

This colliding wave space-time, as  shown in Fig. 1, consists of two
approaching waves, regions II and III, in a flat background, region
IV, and an interaction region, region I. The two waves move in the
direction of two null coordinates $u$ and $v$, and since they have
translational symmetry along the transversal $x$-$y$
planes, the interaction region retains a
two-parameter symmetry group of motions generated by the Killing
vectors $\partial _x$ and $\partial _y$.
The four space-time regions
are separated by the two null wave fronts $u=0$ and $v=0$. Namely, the
boundary between regions I and II is  $\{0\leq u<\pi /2,\; v=0\}$, the
boundary between regions I and III is $\{u=0,\; 0\leq v <\pi /2\}$, and
the boundary of regions II and III with region IV is
$\{u\leq 0,\;v=0\}\cup\{u=0,\;v\leq 0\}$. Region I meets region IV
only at the surface $u=v=0$. The Killing-Cauchy horizon in the region
I corresponds to the hypersurface $u+v=\pi /2$ and plane wave regions
II and III meet such a  Killing-Cauchy horizon
only at ${\cal P}=\{u=\pi /2,\; v=0\}$ and
${\cal P}'=\{u=0,\; v=\pi /2\}$ respectively. 
Observe that plane wave regions II
and III contain a singularity
at $u=\pi /2$, for region II, and $v=\pi /2$, for region III. These
singularities are not curvature singularities but a type of
topological singularity commonly referred to as a folding
singularity \cite{gri91}. This terminology arises from the fact that
the whole singularity $u=\pi /2$ in region II (or $v=\pi /2$ in region III)
must be identified (i.e. ``folded'') with $\cal P$ (or ${\cal P}'$)
(see \cite{dor93} for more details and for a 3-dimensional plot of this
space-time).

\section{Mode propagation}

In this section we will study the interaction of a quantum field with
the colliding plane wave background. In particular we will be
interested in the value of the
quantum field which corresponds to the initial vacuum state in the flat
region, all over the causal past of the collision center (region
$\cal S$ in Fig. 2).
This turns out to be a
geometrical problem which consists in solving the field equation in the
four space-time regions and smoothly joining the distinct solutions.
We start with the field
solution in the flat region, which is chosen to be the usual vacuum state
in Minkowski space-time. This vacuum solution will set a well posed
initial value problem on the null boundaries $\Sigma =\{u=0,\; v\leq
0\}\cup\{u\leq 0,\; v=0\}$, by means of
which a unique solution for the field equation can be found throughout
the space-time, i.e., in the plane wave regions (regions II and III), and in the
interaction region (region I).
However, although it is rather easy to find the solution of the field
equation in regions II and III
which matches smoothly with the boundary conditions, it turns out to be a
difficult problem for the interaction region. The reason is
essentially due to the intrinsic differences between the the geometry of
the plane wave regions and the interaction region. In fact, the plane
wave regions are either conformally flat or type N in the Petrov
classification, but the interaction region can be more generic.
We will refer, from now on, to this problem as {\em the mode
propagation problem}. 

For certain particular examples it has been only partially possible
to solve the mode propagation problem \cite{dor93,dor94,dor96,fei95},
essentially due to the separable properties of the field
equation. There is not, however, an analitical way to solve such a
problem in general, for any colliding plane wave
space-time. Nevertheless, if we are only interested in the collision
center, we do not need 
to solve the mode propagation problem throughout the entire space-time
but only in the  causal past of the collision center 
(i.e region $\cal S$). We will see
in what follows 
that this not only will introduce an important simplification 
but also will allow us to find an
approximation procedure, presumably valid for any colliding plane wave
space-time.

Let us now  solve the field equation in all four space-time
regions.
 We will consider for simplicity a massless scalar field,
which satisfies the usual Klein-Gordon equation,

\bb \Box\phi =0,\label{eq:kG}\ee
and let us take the line element,

\bb ds^2=2{\rm e}^{-M(u,v)}dudv-{\rm e}^{-U(u,v)}\left(
{\rm e}^{V(u,v)}dx^2+{\rm e}^{-V(u,v)}dy^2
\right), \label{eq:dsG}  \ee
which applies globally to the four space-time regions, and where the
functions $U$, $V$ and $M$, can be directly read from
(\ref{eq:ibIV})-(\ref{eq:ibI}). Then, the
field equation can
be separated in a plane-wave form solution for the transversal
coordinates $x$ and $y$, with $k_x$ and $k_y$, respectively, as
separation constants. This plane-wave separation is just a trivial 
consequence of the translational symmetry of the space-time on the planes
spanned by the Killing vectors $\partial _x$ and $\partial _y$. The
field solution is thus,

\bb \phi (u,v,x,y)={\rm e}^{U(u,v)/2}\, f(u,v)\, {\rm e}^{ik_xx+ik_yy},
\label{eq:phiG}\ee
where the function $f(u,v)$ satisfies the following second order
differential equation,

\bb f_{,uv}+\Omega (u,v)\, f=0;\;\;\; \Omega (u,v)=-
{\left({\rm e}^{-U/2}\right)_{,uv}\over {\rm e}^{-U/2}}+{1\over 2}
{{\rm e}^{-M+U}}\left( k_x^2{\rm e}^{-V}+k_y^2{\rm e}^V\right).
\label{eq:f(u,v)}\ee
From now on, we will refer to $\Omega (u,v)$ as the {\em  potential term}.
In the flat region (region IV) this potential term is simply,

\bb \Omega _{\rm IV}(u,v)=L_1L_2(k_x^2+k_y^2), \label{eq:VIV}\ee
where we have used that the functions $U$ and $V$ in (\ref{eq:dsG})
are zero in the flat region, and the funtion $\exp (-M)=2L_1L_2$.
Using (\ref{eq:VIV}), equation (\ref{eq:f(u,v)}) can
be solved as,

\bb f(u,v)={\rm e}^{-i2{\hat k}_+u-i2{\hat k}_-v},\label{eq:fIV}\ee
where $k_\pm$ are two new separation constants with dimensions of
energy, and we define for convenience two dimensionless constants as
${\hat k}_\pm\equiv \sqrt{L_1L_2}\, k_\pm$. These new separation
constants are directly related to the previous ones $k_x$ and $k_y$
by,

\bb 4\, k_+k_-=k_x^2+k_y^2. \label{eq:k+-}\ee
The field solution in region IV reduces, thus, to the usual Minkowski
plane wave solution, i.e,

\bb \phi _k (u,v,x,y)={1\over\sqrt{2k_-(2\pi )^3}}\,
{\rm e}^{-i2{\hat k}_+u-i2{\hat k}_-v+ik_xx+ik_yy}. \label{eq:phiIV}\ee
These modes are well normalized on the null hypersurface
$\Sigma =\,\{(u=0,\,v<0\}\cup\{u<0,\,v=0\}$, which is the boundary
of the plane wave regions II and III with the flat region IV. Even
though $\Sigma$ is a 
null hypersurface,
a well defined scalar product is given by (see 
\cite{dor93} for details),

\bb (\phi _1,\phi _2)=-i\int dx dy\left[
\int _{-\infty} ^0 \left.\left(\phi _1 
{\buildrel\leftrightarrow\over\partial}_u \phi
_2^*\right)\right|_{v=0}\, du +
\int _{-\infty} ^0 \left.\left(\phi _1 
{\buildrel\leftrightarrow\over\partial}_v \phi
_2^*\right)\right|_{u=0}\, dv\right].  \label{eq:scalarproduct}\ee
The modes (\ref{eq:phiIV}) will determine on $\Sigma$ a well 
posed set of boundary
conditions for modes in regions II and III. There, the
potential term in equation (\ref{eq:f(u,v)}) is simply,

\bb \Omega _{i}={1\over 2}{\rm e}^{-M_{i}+U_{i}}
\left(k_x^2{\rm e}^{-V_{i}}+k_y^2{\rm e}^{V_{i}}\right), 
\label{eq:VII/III}\ee
where the label $i={\rm II}$ or $i={\rm III}$ in the functions 
$U$, $V$ and $M$, stands
for their particular values in the plane wave regions II or III.
Then, the solution of equation (\ref{eq:f(u,v)}) in regions II and III with
the boundary conditions imposed by the flat modes (\ref{eq:phiIV}) on
the hypersurface $\Sigma$, can be easily found as,

\bb f(u,v)=\left\{\begin{array}{l}
{\rm e}^{-i2{\hat k}_-v-iA_{\rm II}(u)/(2{\hat k}_-)};\;\;\;\;{\rm in\; region
\; II},\\
\\
{\rm e}^{-i2{\hat k}_+u-iA_{\rm III}(v)/(2{\hat k}_+)};\;\;\;{\rm in\; region
\; III},
\end{array}\right. \label{eq:fII/III}\ee
where the generic function $A_i(\zeta )$, with $i=$II, III is given by,

\bb A_i(\zeta )=\int_0^\zeta  d\zeta '\, {1\over 2}
{\rm e}^{-M_i(\zeta ')+U_i(\zeta ')}
\left({\rm e}^{-V_i(\zeta ')}k_x^2+{\rm e}^{V_i(\zeta ')}k_y^2\right).
\label{eq:AII/III}\ee
Therefore, the well normalized ``in'' modes in regions II and III are,

\bb \phi (u,v,x,y)={1\over\sqrt{2k_-(2\pi)^3}}\,{\rm e}^{ik_xx+ik_yy}\,
\left\{\begin{array}{l}
\displaystyle
{1\over\cos u}{\rm e}^{-i2{\hat k}_-v-iA^{}_{\rm II}(u)/(2{\hat k}_-)};
\;\;\;\;{\rm in\; region\; II},\\
\\
\displaystyle
{1\over\cos v}{\rm e}^{-i2{\hat k}_+u-iA^{}_{\rm III}(v)/(2{\hat k}_+)};
\;\;\;{\rm in\; region\; III},
\end{array}\right. \label{eq:phiII/III}\ee
where, using the notation $k_1=\sqrt{L_1L_2}\, k_x$, $k_2=\sqrt{L_1L_2}\, k_y$,

\bb A^{}_i(\zeta )=k_1^2\,\left[
{(1+\sin \zeta )^2(9-\sin \zeta )\over 2\cos \zeta }
+{15\over 2}\cos \zeta -{15\over 2}\zeta -12
\right]+k_2^2\,\tan \zeta , \label{eq:Aib}\ee

Now the Cauchy problem is well posed on the boundaries $\{u=0,\,0\leq
v<\pi /2\}$
and $\{0\leq u<\pi /2,\, v=0\}$ between plane wave regions II and III and
the interaction region. Notice that the initial modes
(\ref{eq:phiII/III}) are well normalized on  the boundary $\Sigma$
between the flat region and the plane wave regions, and this means,
from general grounds, that they remain well normalized on the boundary
between the plane waves and the interaction region.

We now have to solve equation (\ref{eq:f(u,v)}) in region I with the
boundary conditions imposed by (\ref{eq:phiII/III}) on the lines
$\Sigma _{\rm I}=\,\{u=0,\, 0\leq v<\pi /2\}\cup\{0\leq u<\pi /2,\, v=0\}$. These
lines are characteristic lines for the partial differential equation 
(\ref{eq:f(u,v)}), and therefore the only independent boundary
conditions, i.e. the Cauchy data, are the initial values of the
function $f(u,v)$ on them (the normal derivatives of
the function on the characteristics, which are usually part of the Cauchy
data, are determined by the values of the function $f(u,v)$ itself
\cite{gara64}).
In fact, we only need to find the solution of equation
(\ref{eq:f(u,v)}) in the neighborhood of the collision center and
from general grounds in
partial differential equation theory (see \cite{gara64} for details) this
solution will be determined only by a
subset of the Cauchy data (see Fig. 2). We will see in what follows
that since that subset of Cauchy data includes only smooth
non-singular
functions, the mode propagation problem can be suitably approximated.
We start with the change of coordinates,

\bb {t} =u+v,\;\;\; {z}=v-u, \label{eq:xieta}\ee
in  equation (\ref{eq:f(u,v)}) and we obtain,

\bb f_{,{t}{t}}-f_{,{z}{z}}+\Omega ({t} ,{z})f=0,  \label{eq:fxieta}\ee
where the potential term is given by,

\bb \Omega ({t} ,{z})={(1+\sin{t})^4k_1^2+(\sin ^2{t})/4\over\cos ^2{t}}+
{k_2^2-(\sin ^2{z})/4\over\cos ^2{z}}. \label{eq:Vxieta}\ee
The coordinate singularity (Killing-Cauchy horizon)
occurs at ${t} =u+v=\pi /2$. The lines 
$\Sigma _{\rm I}=\,\{u=0,\,0\leq v<\pi /2\}\cup\{0\leq u<\pi /2,\, v=0\}$,
where the Cauchy data is imposed  by (\ref{eq:phiII/III}), are
characteristics for the equation (\ref{eq:f(u,v)}), or (\ref{eq:fxieta}). 
The collision center is determined by the simple condition $u=v$, and 
thus the subset of Cauchy data that affects the neighborhood of
$u=v$ lies on the lines,
${\bar\Sigma}_{\rm I}=\,\{u=0,\,0\leq v<\pi /4\}\cup\{0\leq u<\pi /4,\,
v=0\}$. Region $\cal S$ in  Fig. 2 is the causal future of this
Cauchy data (or equivalently, the causal past of the colision center).

However, the behaviour of the variables ${t}$ and ${z}$ in equation
(\ref{eq:fxieta}) in region $\cal S$ is
very different. Since ${t}$ runs from $0$ to $\pi /2$ and ${z}$ runs
from $-\pi /4$ to $\pi /4$ in this region, the potential term 
(\ref{eq:Vxieta}) blows up as coordinate ${t}$ goes to $\pi /2$, but
is perfectly smooth over the entire range of coordinate
${z}$. This fact suggests that in the whole region $\cal S$
the physical results that we may expect are directly
related to the coordinate ${t}$ and we may not expect any physically
remarkable change if we take ${z} =0$ in equation (\ref{eq:fxieta}). 
However, if
we want to be consistent with such an approximation, we must also
modify the boundary conditions that lie on the line segments
${\bar\Sigma}_{\rm I}$. Since on the boundary ${\bar\Sigma}_{\rm I}$
we have that 
${t} =\pm{z}$ and coordinate ${t}$ runs from $-\pi /4$ to $\pi /4$,
we must also take ${t} ={z} =0$. This means that the boundary
conditions on ${\bar\Sigma}_{\rm I}$, given by (\ref{eq:phiII/III}), reduce
in such an approximation
to the flat boundary conditions (\ref{eq:phiIV}). Therefore we
change the mode propagation problem for the colliding wave space-time
into a rather simpler problem, which is
clear from Fig. 3, and which requires only that we find a solution
to equation (\ref{eq:fxieta}) with initial conditions given by the
Minkowski flat modes (\ref{eq:phiIV}) below the hypersurface
$\{{t} =0,\, -\pi /4 <{z}<\pi /4\}$. In fact, using such a
simplification we can explicitly eliminate the
dependence of equation (\ref{eq:fxieta}) on
coordinate ${z}$, with ${z} =0$, 
by taking a ${z}$-plane wave solution, i.e.,
$f({t} ,{z} )={g}({t} )\, {\rm e}^{ik_3{z}}$,
where $k_3$ is a new separation constant.

Observe that we have changed a difficult problem of two-variable
partial differential equations into a rather easy one-dimensional
Schr\"odinger-type problem. Recall, however,  that  
none of the discussion above is
applicable when a solution of equation
(\ref{eq:fxieta}) in the neighborhood of the folding singularities
${\cal P}$ and ${\cal P}'$ is required. This is not
only because in that case  both coordinates ${t} $ 
and ${z}$ take values near $\pi
/2$ and thus the potential term (\ref{eq:Vxieta}) is unbounded as
$z\rightarrow\pi /2$, but
also because the boundary conditions (\ref{eq:phiII/III}) are also
unbounded as the folding singularities at ${t} =\pm{z} =\pi /2$ are approached.
In that case the mode propagation problem is much more complicated and
a more detailed discussion is required (see
\cite{dor93,dor94,dor96,fei95}) for details).

In order to solve this Schr\"odinger-type problem, rather than relying
on the discussed approximations for the exact field equation
(\ref{eq:fxieta}), we will rewrite a new field equation using an
adequate approximation for the line element throughout the entire
causal past of the collision center.
This new approach, which  may seem redundant in the case of
the {\em particle production problem}, is absolutely necessary when
the renormalization of the stress-energy tensor is
discussed. This is essentially because the process of renormalization involves
the subtraction of the infinite divergences that arise from the
formal definition of the stress-energy tensor, and these divergences
can be expressed as entirely geometric terms, which are independent
of any possible 
approximations in the field equation. This means that in order to
recover the geometric divergences in the stress-energy tensor,  any
approximation in the field equation solutions must be 
related to a suitable  approximation in the space-time geometry.

Approximating the field equation (\ref{eq:fxieta}) in the
causal past of the collision center by taking ${z} =0$ is essentially
equivalent to changing 
the line element, in the causal
past of the collision center, by a related line element
obtained directly from the 
original (\ref{eq:ibI}) by setting ${z} =0$, i.e.,

\bb
d{\hat s}^2_{\rm I}=L_1L_2(1+\sin{t})^2\left(d{t} ^2-d{z} ^2\right)
-{1-\sin{t}\over 1+\sin{t}}\, dx^2-(1+\sin{t})^2\, dy^2.
\label{eq:dshatIb}
\ee
We will suppose that the line element (\ref{eq:dshatIb})
applies all over the causal past of the collision center, not
only in the interaction region but also through the plane wave regions
II and III in the sense of Fig. 3.
The plane wave collision starts at $t=0$ but
to avoid smoothness problems derived from such an
approximation, we will suppose that (\ref{eq:dshatIb}) applies
exactly on a range $\epsilon <{t}<\pi /2$, for a certain
$\epsilon >0$. In the range
$0\leq {t}\leq\epsilon$, as described
below,  we will interpolate a line element 
which smoothly matches with the flat space at $t=0$.
Nevertheless, the details of this matching will not affect the main physical
features.

The
exact field equation for this approximate space-time is,

\bb
\left(\Box +\xi  R\right)\phi =0,
\label{eq:Appfieldeq}\ee
where it is necessary to consider a coupling curvature term in the field
equation because, although the exact space-time is a vacuum
solution, we have a bounded nonzero value for
$R$ in the approximated space-time.
In order to solve this new field equation, we start rewriting  the line
element (\ref{eq:dshatIb}) in the following general way,

\bb
ds^2=(f_1f_2f_3)\, d{{t} ^*}^2-\left(f_1f_2\over f_3\right)\, d{z} ^2
-\left(f_2f_3\over f_1\right)\, dx^2-
\left(f_1f_3\over f_2\right)\, dy^2,
\label{eq:dsgen}
\ee
where the $f_i$ are functions of coordinate ${t}$ alone, which
for values of $0<\epsilon <{t}<\pi /2$, can be
straightforwardly determined by direct comparison with
(\ref{eq:dshatIb}) as $f_1({t})=\sqrt{L_1L_2}\, (1+\sin{t})^2$,
$f_2({t})=\sqrt{L_1L_2}\, \cos{t}$, $f_3({t})=\cos{t}$.
For values ${t}\leq 0$ we take $f_1({t})=f_2({t})=\sqrt{L_1L_2}$, 
$f_3({t})=1$, which correspond to their values in flat
space. Finally,  in
the interval $0\leq{t}\leq\epsilon$, we smoothly interpolate each  $f_i({t})$
($i=1,2,3$) between these values.
Also,
in order to prevent singularities in the field
equation, we conveniently reparametrize coordinate ${t}$, by ${t}
^*({t})$, as follows,

\bb
{d{t} ^*\over d{t}}={1\over f_3({t})}.
\label{eq:dxi*dxi}\ee
Now, we use the following ansatz for the field solutions,

\bb
\phi _k=h({t} ^*)\,{\rm e}^{ik_xx+ik_yy+ik_{z}{z}},
\label{eq:ansatz}\ee
where the plane wave factor in coordinates $x$, $y$ is related to the
translational symmetry of the space-time along  the transversal directions
$x$, $y$, and the plane wave factor in
coordinate ${z}$ is just a consequence of our approximation. 
Then equation (\ref{eq:Appfieldeq})
directly leads to the following Schr\"odinger-like differential
equation for the function $h({t} ^*)$,

\bb
h_{,{t} ^*{t} ^*}+\omega ^2({t} )\, h=0,\;\;\;\;
V({t})\equiv\omega ^2({t} )=f_0^2({t} )+f_1^2({t} )\, 
k_x^2+f_2^2({t} )\, k_y^2+
f_3^2({t} )\, k_z^2,
\label{eq:heq}\ee
where the function $f_0({t})$ stands for,

\bb
f_0^2({t} )=\left[f_1({t} )f_2({t} )f_3({t} )\right]\,\xi  R.
\label{eq:omegafi}\ee
Such differential equation can be WKB solved, essentially because the
short wavelength condition holds, i.e. $\omega
^{-1}d/d{t}^*\ln\omega\ll 1$. Observe that this condition reduces to
$(d{t}/d{t}^*)\, dV/d{t}\ll 2\,\omega ^3$, which becomes
particularly accurate when the Killing-Cauchy horizon is approached
since in that case $d{t}/d{t}^*=f_3({t})\rightarrow 0$. Therefore, the mode
solutions
$\phi _k$ which reduce to the flat mode solutions in the region prior
to the arrival of the waves, are

\bb
\phi _k={{\hat\omega}^{1/2}\over\sqrt{(2\pi)^32k_-W({t} )}}
{\rm e}^{ik_xx+ik_yy+ik_3{z}-i\int ^{{t} ^*}W(\zeta )d\zeta ^*},
\label{eq:phifi}\ee
where we denote ${\hat\omega}^2=k_1^2+k_2^2+k_3^2$
with $k_1=\sqrt{L_1L_2}\, k_x$, $k_2=\sqrt{L_1L_2}\, k_y$,
$k_3=k_z$ and where $W({t} )$ stands for an
adiabatic series in powers of the time-dependent frequency 
$\omega ({t} )$ of the 
modes and its derivatives. Up to adiabatic order four (i.e. up
to terms involving four derivatives of $\omega ({t})$) 
$W({t})$ it is given by,

\bb W({t} )=\omega  +{A_2\over\omega ^3} +{B_2\over\omega ^5} 
+{A_4\over\omega ^5}
 +{B_4\over\omega ^7} +{C_4\over\omega ^9} +{D_4\over\omega 
^{11}},\label{eq:Wxi}\ee
where, using the notation ${\dot V}\equiv dV/d{t}^*$,
 
\bb A_2=-{{\ddot V}\over 8},\;\;\;\; B_2={5\over 32}\,{{\dot 
V}^2},\label{eq:An}\ee
 
$$ A_4={{\ddddotV}\over 32},\;\;\;\; B_4=-
{28\,{\dot V}\,{\dddotV}+19\,{\ddot
V}^2\over 128},\;\;\;\; C_4={221\over 258}\,{{\dot V}^2\,{\ddot V}},\;\;\;\;
D_4=-{1105\over
2048}\,{{\dot V}^4},$$
and 
$A_n$, $B_n$, ... denote the $n$ adiabatic terms in
$W({t})$.
Up to adiabatic order zero it is simply
$W({t} )=\omega({t} )$. Observe
that in the flat region prior to the arrival of the waves we have
$W({t} )={\hat\omega}=\left(k_1^2+k_2^2+k_3^2\right)^{1/2}$. Observe 
also that since
$f_3=1$ in this flat region, we can use (\ref{eq:dxi*dxi}) to set
$ {t} ^*={t}$, where without loss of generality we choose the
origin ${t}^*=0$ at ${t=0}$.
Therefore, the mode solutions (\ref{eq:phifi}) in the flat region reduce to,

$$
\phi _k^{{\rm IV}}={1\over\sqrt{(2\pi)^32k_-}}{\rm
e}^{ik_xx+ik_yy+ik_{z}{z} -i{\hat\omega}{t}},
$$
which indeed are the flat mode solutions defined in (\ref{eq:phiIV}),
recalling that the new separation constant $k_{z}=k_3$ is related to the
original $k_\pm$ by the ordinary null momentum relations, i.e.,

\bb
{\hat\omega}={\hat k}_++{\hat k}_-,\;\;{k_{z}}={\hat k}_+-{\hat k}_-.
\label{eq:nullmom}\ee

It is important to understand that we are constructing a set of mode
solutions as an adiabatic series in terms of derivatives of the
frequency $\omega ({t})$ in the differential equation
(\ref{eq:heq}). This procecure is similar but not equivalent to the
construction of an {\em adiabatic vacuum state} where the field modes
are expanded as an adiabatic series in terms of the derivatives of the
metric coefficients (see for example \cite{bir82} for details). In
fact, observe for instance that the term $f^2_0({t})$ in (\ref{eq:heq})
involves two derivatives of the metric since it is directly related to
the curvature scalar and therefore it would be an adiabatic order two
term for an eventual adiabatic vacuum construction, but it is simply
an order zero term in our adiabatic series in derivatives of 
$\omega ({t})$.

\section{``Point splitting" regularization technique}
 
In this section we briefly review, for the computational purposes of the 
following sections, the
``point-splitting" regularization technique to calculate the expectation value 
of the stress-energy tensor of a
scalar quantum field in some physical state. The stress-energy tensor of the 
field may be obtained by functional
derivation of the action for the scalar field with respect to the metric. When 
the field is massless, it is
\cite{bir82}
 
\bba T_{\mu\nu}&=&(1-2\xi)\,\phi _{;\mu}\,\phi _{;\nu}+(2\xi-{1\over 
2})\,g_{\mu\nu}\,g^{\alpha\beta}\,\phi
_{;\alpha}\,\phi _{;\beta}-2\xi\,\phi _{;\mu\nu}\,\phi +{1\over 2}\,\xi\, 
g_{\mu\nu} \,
\phi{\nabla}^{\alpha}{\nabla}_{\alpha}\phi\nonumber\\
& &-\xi\,\left(R_{\mu\nu}-{1\over 2}R\, g_{\mu\nu}+
{3\over 2}\xi R\, g_{\mu\nu}\right)\phi ^2
 ,\label{eq:TS1}\eea
where $\xi$ is the coupling parameter of the field to the curvature.
 
To quantize, the
field $\phi$ is promoted into a field operator acting over a given Hilbert 
space $H_{\phi}$
\cite{bir82,wal94}, 
$\phi (x)=\sum _k a_k u_k(x)+a^{\dag}_k u^*_k(x)$,
where $a^{\dag}_k$, $a_k$ are the standard creation and annihilation operators 
and $\{u_k(x)\}$ is a complete
and orthornormal set of solutions of the Klein-Gordon equation 
(\ref{eq:Appfieldeq}). 
Mathematically the field
operator $\phi (x)$ is a point distribution, therefore, the quantum version of 
the stress-energy tensor
(\ref{eq:TS1}) is mathematically pathological because it is quadratic in the
field and its derivatives. One possible way to give  sense  to that expression 
is
to note that the formula (\ref{eq:TS1}) can be formally recovered as,
 
\bb \langle T_{\mu\nu}(x)\rangle =\lim _{x\rightarrow x'}\, {\cal 
D}_{\mu\nu}G^{(1)}(x,x'), \label{eq:limDT}\ee
where $G^{(1)}(x,x')$ is a Green's function of the field equation defined as the 
vacuum expectation value of the
anticommutator of the field, and called the {\it Hadamard function}, 
 
\bb G^{(1)}(x,x')=\langle\{\phi (x),\phi (x')\}\rangle =\sum _k 
\left\{u_{k}(x)\, u^*_{k}(x')+u_{k}(x')\,
u^*_{k}(x)\right\}. \label{eq:defHadamard}\ee
As a product of
distributions at different points this is mathematically well defined. The 
differential operator ${\cal
D}_{\mu\nu}$ is given in our case by,
 
\bba{\cal D}_{\mu\nu}&=&
\left(1-2\,\xi\right)\,{1\over 4}\,\left(\nabla _{\mu '}\nabla _{\nu}+\nabla
_{\nu'}\nabla_{\mu}\right)
+  (2\,\xi-{1\over 2})\, g_{\mu\nu}\,{1\over 4}\,\left(
\nabla _{\alpha '}\nabla ^{\alpha}+
\nabla _{\alpha }\nabla ^{\alpha '}\right)-\nonumber\\
& &{\xi\over 2}\,\left(\nabla _{\mu }\nabla _{\nu}+\nabla
_{\mu'}\nabla_{\nu '}\right)
+   g_{\mu\nu}\,{\xi\over 8}\,\left(
\nabla _{\alpha  }\nabla ^{\alpha  }+
\nabla _{\alpha '}\nabla ^{\alpha '}\right)-\nonumber\\
& &{\xi\over 2}\,\left(R_{\mu\nu}-{1\over 2}R\, g_{\mu\nu}+
{3\over 2}\xi R\, g_{\mu\nu}\right).\label{eq:Dopdif}\eea
However, the above differential operation and its limit have no immediate 
covariant meaning because
$G^{(1)}(x,x')$ is not an ordinary function but a {\it biscalar} and the 
differential operator ${\cal
D}_{\mu\nu}$ is {\it nonlocal}; thus we need to deal with the nonlocal 
formalism of {\it bitensors} (see, for example \cite{dew60,chr76} or the
Appendix B of reference \cite{dor96} for a review on
this subject).
 
The above procedure still leads to a divergent quantity since we know that 
even in flat space-time
$G^{(1)}(x,x')$ has a short-distance singularity and that a ``vacuum" 
subtraction has to be performed
to $G^{(1)}(x,x')$ in order to obtain a regularized value. To regularize we 
assume that
$G^{(1)}(x,x')$,  has a short-distance singular structure given by
 
\bb S(x,x')={2\over (4\pi)^2}\Delta ^{1/2}(x,x')\left[-{2\over\sigma
(x,x')}+v(x,x')\ln\sigma(x,x')+w(x,x')\right],      \label{eq:Hada}\ee
where $\sigma (x,x')=(1/2)\, s^2(x,x')$ is the {\it geodetic biscalar}
(being $s(x,x')$ the proper distance between $x$ and $x'$ along a non-null
geodesic connecting them),
$\Delta$ is the {\it Van Vleck-Morette} determinant \cite{dew60}, which is 
singularity free in the
coincidence limit, and where $v(x,x')$ and $w(x,x')$ are biscalars with a 
well-defined coincidence limit for
which we assume the following covariant expansions,
 
\bb v(x,x')=\sum _{l=0}^{\infty}v_l(x,x')\sigma ^l(x,x'),\;\;\;\; w(x,x')=\sum 
_{l=0}^{\infty}w_l(x,x')\sigma
^l(x,x').                      \label{eq:Hvw}\ee 
A Green's function expressed in this form is usually called an
{\it elementary Hadamard solution}, the name of which comes from the work of 
Hadamard on the singular structure
for elliptic and hyperbolic second order differential equations. Note, 
however, that this Hadamard
singular structure is not a general feature of any Green's function of the 
Klein-Gordon equation. In other words,
although for an extensive range of space-time and vacuum states, the vacuum 
expectation value of the
anticommutator of the field,
$G^{(1)}(x,x')$, has this singular form, this is not a general property. 
However a theorem
states that if
$G^{(1)}(x,x')$ has the singular structure of an elementary Hadamard solution  
in a neighbourhood of a Cauchy
surface of an arbitrary hyperbolic space-time, then it has this structure 
everywhere \cite{kay91,ful78}.
As a corollary of this theorem, $G^{(1)}(x,x')$ has this singular structure if 
the space-time is flat to the past
of a space-time Cauchy surface, as is the case of our colliding plane wave 
space-time. This and other
considerations led to a proposal by Wald \cite{wal94} that any physically
reasonable quantum state must be a {\it Hadamard state}, that is to say, a 
state for which $G^{(1)}(x,x')$
takes the short-distance singular structure of an elementary Hadamard solution.
 
The coefficients $v_i(x,x')$ and $w_i(x,x')$ can
be directly obtained by substitution in the differential equation,
$(\Box _x+\xi R) S(x,x')=0$.
Recursion relations for $v_i(x,x')$ and $w_i(x,x')$ are then obtained 
\cite{dew60,adl77}. These relations 
uniquely determine
all the $v_i(x,x')$ coefficients  but the coefficients
$w_i(x,x')$ can be written in terms of an arbitrary term $w_0(x,x')$. Up to 
order $\sigma$,
$v(x,x')$ is given by
 
\bb v(x,x')=a_1(x,x')-{1\over 2}\, a_2(x,x')\,\sigma +\cdots ,   
\label{eq:va1a2}\ee
where  $a_1(x,x')$ and $a_2(x,x')$ are the
Schwinger-DeWitt coefficients \cite{bir82} whose midpoint expansion
are given,  up to the order 
required for the regularization calculations, by
 
\bb a_1(x,x')=a_1^{(0)}+{a_1^{(2)}}_{{\bar\mu}{\bar\nu}}\,
\sigma ^{\bar\mu}\sigma ^{\bar\nu}+\cdots ,\;\;\;\;
a_2(x,x')=a_2^{(0)}+\cdots ,
\label{eq:a12}\ee
and the terms $a_1^{(0)}$, ${a_1^{(2)}}_{{\bar\mu}{\bar\nu}}$ and
$a_2^{(0)}$ are written in Apendix A in terms of geometrical quantities.
 
Only the coefficients $v_i(x,x')$ are related to the singular structure of 
$G^{(1)}(x,x')$ in
the coincidence limit, and they are uniquely determined by the space-time 
geometry. This means that
given any two Hadamard elementary solutions in a certain space-time geometry, 
both have the same singularity
structure in the coincidence limit; therefore given two vacuum Hadamard 
states, $| 0\rangle$ and
$|{\overline 0}\rangle$, $G^{(1)}(x,x')=\langle 0|\{\phi (x),\phi 
(x')\}|0\rangle$ and 
${\overline G}^{(1)}(x,x')=\langle {\overline 0}|\{\phi (x),\phi 
(x')\}|{\overline 0}\rangle$, they have the
same singular structure. Their finite parts, however, may differ because the 
two vacuum states are related to
different boundary conditions, which are global space-time features. 
Mathematically this comes from the fact that
the term $w_0(x,x')$ in the elementary Hadamard solution is totally arbitrary; 
fixing $w_0(x,x')$ we fix
a particular boundary condition. This suggests a possible renormalization 
procedure
\cite{wal94,wal78,adl77}: we can eliminate
the non-physical divergences of any $G^{(1)}(x,x')$ without alterations in the 
particular physical boundary
conditions by subtracting an elementary Hadamard solution with the particular 
value $w_0(x,x')=0$, which
is the value that
corresponds to the flat space case. In other words, we define the following 
regularized biscalar,
 
\bb G_B^{(1)}(x,x')=G^{(1)}(x,x')-\left.S(x,x')\right|_{w_0=0}.\label{eq:GB}\ee
This particular value for the elementary Hadamard solution, which we
may refer to as the locally constructed Hadamard function as opposed to
its non-local counterpart $G^{(1)}$, can be easily calculated. In
fact, setting
$w_0(x,x')=0$, the biscalar $w(x,x')$ reduces to \cite{adl77},

\bb w(x,x')=-{3\over 4}a_2^{(0)}\sigma +\cdots .\label{eq:wa2}\ee
and using the following midpoint expansion for the Van Vleck-Morette
determinant we find,

\bb 
\Delta ^{1/2}(x,x')=1
+{\Delta ^{(2)}}_{{\bar\mu}{\bar\nu}}\,\sigma ^{\bar\mu}\sigma ^{\bar\nu}
+{\Delta ^{(4)}}_{{\bar\mu}{\bar\nu}{\bar\rho}{\bar\tau}}
\,\sigma ^{\bar\mu}\sigma ^{\bar\nu}\sigma ^{\bar\rho}\sigma ^{\bar\tau},
\label{eq:VV-M}\ee 
where the coefficients ${\Delta ^{(2)}}_{{\bar\mu}{\bar\nu}}$,
${\Delta ^{(4)}}_{{\bar\mu}{\bar\nu}{\bar\rho}{\bar\tau}}$ are written
in Appendix A. Then
the midpoint expansion for the locally constructed Hadamard
function (\ref{eq:Hada}) reduces to,

\bba
\left. S(x,x')\right|_{w_0=0}&=&{1\over 8\pi ^2}\left\{
-{2\over\sigma}
-2{\Delta ^{(2)}}_{{\bar\mu}{\bar\nu}}\,{\sigma ^{\bar\mu}\sigma
^{\bar\nu}\over\sigma}
-2{\Delta ^{(4)}}_{{\bar\mu}{\bar\nu}{\bar\rho}{\bar\tau}}
\,{\sigma ^{\bar\mu}\sigma ^{\bar\nu}\sigma ^{\bar\rho}\sigma
^{\bar\tau}\over\sigma}
 -a_1^{(0)}\,\ln (\mu ^{-2}\sigma ) 
\right.\nonumber
\\
& & \left.
-\left[
\left(
a_1^{(0)}{\Delta ^{(2)}}_{{\bar\mu}{\bar\nu}}
+{a_1 ^{(2)}}_{{\bar\mu}{\bar\nu}}
\right)\sigma ^{\bar\mu}\sigma ^{\bar\nu}
-{1\over 2}a_2^{(0)}\sigma
\right]\,\ln (\mu ^{-2}\sigma )
-{3\over 4}a_2^{(0)}\sigma
\right\},
\label{eq:Hada1}\eea
where we have included an arbitrary length parameter $\mu$ in the
logarithmic term which, as we will mention below, is related to
the two-parameter ambiguity of the point-splitting regularization scheme.

Then by means of $G_B^{(1)}(x,x')$ we can construct a 
$\left<T_{\mu\nu}^B\right>$ by differentiation with respect to the
nonlocal operator (\ref{eq:Dopdif}). 
This regularization procedure, however, fails to give a covariantly conserved 
stress-energy tensor essentially because the locally constructed
Hadamard function (\ref{eq:Hada1}) is not in
general symmetric on the endpoints $x$ and $x'$ (i.e. it satisfies the
field equation at the point $x$ but fails to satisfy it at $x'$). In fact,
it can be seen  \cite{wal78} that for a massless conformal
scalar field (i.e.
$\xi =1/6$),
 
\bb \nabla ^{\nu} \langle T_{\mu\nu}^B\rangle
={1\over 4}\,\lim _{x\rightarrow x'}\nabla _{\mu}\,
\left(\Box _{x'}+{1\over 6}\, R(x')\right)\,
G^B(x,x')={1\over 64\pi ^2}\, \nabla _{\mu}\, a^{(0)}_2(x),
  \label{eq:anomalia}\ee 
where $a^{(0)}_2(x)$ is the coincidence limit of the Schwinger-DeWitt
coefficient $a_2(x,x')$ given in Appendix A. 
Thus to ensure covariant conservation, we must introduce an additional 
prescription:
 
\bb \langle T_{\mu\nu}(x)\rangle =\langle T_{\mu\nu}^B (x)\rangle
- {a^{(0)}_2(x)\over 64\pi ^2}\, g_{\mu\nu}. \label{eq:GBT}\ee
Note that this last term is responsible for the trace anomaly in the conformal 
coupling case, because even though
$\langle T_{\mu\nu}^B (x)\rangle$ has null trace when $\xi =1/6$, the trace of 
$\langle T_{\mu\nu}(x)\rangle$
is given by
$\langle T^{\mu}_{\mu}\rangle =
- {a^{(0)}_2(x)/(16\pi ^2)}$.
Recall that for scalar fields with non-null mass we may regularize
the Hadamard 
function by subtracting a truncated DeWitt-Schwinger series (see
\cite{bir82} for details). This gives a symmetric regularized Hadamard
function by means of which an automatically conserved stress-energy
tensor can be obtained. Nevertheless, the DeWitt-Schwinger series
is not well defined in the massless limit and therefore it is not
suited to our case.
 
The regularization prescription just given in (\ref{eq:Hada1})
satisfies the well known four  
Wald's axioms
\cite{wal94,wal76,wal77-78b,chr75}, a set of properties that any physically 
reasonable expectation value of the
stress-energy tensor of a quantum field should satisfy. There is still an 
ambiguity in this prescription since
two independent conserved local curvature terms, which are quadratic in the 
curvature, can be added to this
stress-energy tensor. In particular, the $\mu$-parameter ambiguity in 
(\ref{eq:Hada1}) is a consequence of this (see \cite{wal94} for details).
Such a two-parameter ambiguity, however, cannot be 
resolved within the limits of the
semiclassical theory, it may be resolved in a complete quantum theory
of gravity 
\cite{wal94}. Note, however, that in some sense this ambiguity does not affect 
the knowledge of the matter
distribution because a tensor of this kind belongs properly to the left hand 
side of Einstein equations, i.e.
to the geometry rather than to the matter distribution.

\section{Hadamard function in the interaction region}
 
Here we calculate the Hadamard function $G^{(1)}(x,x')$ 
in the interaction region for the initial vacuum state
defined by the modes $\phi _k$, (\ref{eq:phifi}). The 
Hadamard function can be written as,
 
\bb G^{(1)}(x,x')=\sum _k\phi
_k(x)\,\phi ^*_k(x')\; +{c.c.}    
\label{eq:G(1)}\ee
Note that solutions $\phi _k$ contain the function $h({t} ^*)$, which
cannot be calculated analytically but may be approximated
up to any adiabatic order as described in (\ref{eq:Wxi})-(\ref{eq:An}). 
This means that we have the 
inherent ambiguity of where to
cut the adiabatic series.
However, observe from (\ref{eq:dxi*dxi}) and (\ref{eq:omegafi})
that since $dt/dt^*\rightarrow 0$ and
$V({t})=\omega ^2({t})\rightarrow 16\, k_1^2$ 
towards the horizon, the adiabatic series (\ref{eq:Wxi})
reduces to $W\simeq\omega$ near the horizon. This means that we could cut the 
adiabatic series (\ref{eq:Wxi}) at order zero if we were
interested in a calculation near the horizon. However, this is only
partially true. In fact, it would be true if we were only interested in the 
particle production problem (see \cite{dor93}) for details) but it is not
sufficient for the calculation of the vacuum expectation value of the
stress-energy tensor.
This is because
$G^{(1)}$ calculated with $h({t}^*)$
at order zero does not reproduce the short-distance singular 
structure of a Hadamard elementary
solution (\ref{eq:Hada}) in the coincidence limit $x\rightarrow x'$. The 
smallest adiabatic
order for the function $h({t}^*)$ 
which we need to recover the singular structure of $G^{(1)}$ is order 
four, basically  because our adiabatic construction af the mode
solutions is similar (but not equivalent) to an {\em adiabatic vacuum
state} (see \cite{bir82} for details).
 
Although expanding the function $h({t}^*)$ in (\ref{eq:G(1)}) up to 
adiabatic order four will give an accurate value for
the stress-energy tensor near the horizon, it will also give a suitable
approximate value for this tensor 
all over the causal past of the collision center
(region $\cal S$ in Fig. 3). The reason is that even though the 
short-wavelength condition, i.e.
$\omega ^{-1}d/d{t}^*\ln\omega\ll 1$, is particularly accurate near the
horizon it also holds throughout region $\cal S$.

In the mode sum (\ref{eq:G(1)}) we use the shortened notation 
$\sum 
_k\equiv 
\int ^{\infty}_{0}{dk_-/ k_-}\,
\int ^{\infty}_{-\infty}dk_x\,
\int ^{\infty}_{-\infty}dk_y$ or equivalently
$\sum _k\equiv
(L_1L_2)^{-1}
\int ^{\infty}_{-\infty}dk_1\,
\int ^{\infty}_{-\infty}dk_2\,
\int ^{\infty}_{-\infty}{dk_3/ {\hat\omega}}$,
where the change of variables
(\ref{eq:nullmom}) and the usual notation $k_1=\sqrt{L_1L_2}\, k_x$, 
$k_2=\sqrt{L_1L_2}\, k_y$ have been used.
Therefore we have, 
            
\bba  G^{(1)}(x,x')&=& {1\over 2(2\pi)^3\, L_1L_2}\,
\int ^{\infty}_{-\infty}\int ^{\infty}_{-\infty}\int
^{\infty}_{-\infty}
{dk_1\, dk_2\, dk_3\over \sqrt{W({t})W({t} ')}}\,\times \nonumber\\
& &{\rm e}^{-i\int _{{t^*} '}^{{t^*}}W({\zeta})d{\zeta}^*+ik_x
(x-x')+ik_y (y-y')+ 
ik_{z} ({z} -{z} ')}\;\; +c.c.                         
\label{eq:G(1)2}\eea
 
Let us start now with the point splitting procedure. We assume that the points 
$x$ and $x'$ are connected by a
non-null geodesic in such a way that they are at the same proper distance 
$\epsilon$ from a third midpoint
$\bar x$. We parametrize the geodesic by its proper distance $\tau$ and 
with abuse of notation we denote 
the end points by
$x$ and $x'$, which should not be confused with the third
component of $({t} ,\;{z} ,\; x,\; y)$. Then we expand,
 
\bb x^{\mu}=x^{\mu}(\tau)=x^{\mu}({\bar\tau}+\epsilon)=x^{\mu}_0+\epsilon\, 
x^{\mu}_1+ 
{\epsilon ^2\over 2!}\, x^{\mu}_2+\cdots , \label{eq:xm}\ee 
\bb x'^{\mu }=x^{\mu }(\tau ')=x^{\mu }({\bar\tau}-\epsilon)=x^{\mu 
}_0-\epsilon\, x^{\mu }_1+ 
{\epsilon ^2\over 2!}\, x^{\mu }_2-\cdots , \label{eq:xm'}\ee 
where we have defined,
 
\bb x^{\mu}_0\equiv x^{\mu}({\bar\tau}),\;\;\; x^{\mu}_1\equiv \left.{d
x^{\mu}\over d\tau}\right|_{\bar\tau}, \;\;\; x^{\mu}_2\equiv\left.{d^2
x^{\mu}\over d\tau ^2}\right|_{\bar\tau}, \cdots ,                  
\label{eq:xn}\ee
i.e. the subscript of a given coordinate indicates the number of derivatives 
with respect to $\tau$ at the point
$\bar\tau$. The geodetic biscalar can then be written as

\bb
\sigma = 2\epsilon ^2\Sigma ,\;\;\; 
{\sigma}^{\bar\mu}=2\epsilon x_1^{\bar\mu},\;\;\; 
x_1^{\bar\mu}{x_1}_{\bar\mu}=\Sigma ,
\label{eq:sigeps}\ee 
with $\Sigma =+1$ if $x_1^{\bar\mu}$ is timelike and 
$\Sigma =-1$ if $x_1^{\bar\mu}$ is spacelike and where 
${\sigma}^{\bar\mu}$ is the 
geodesic tangent vector at the midpoint $\bar x$ with modulus the
proper distance on the geodesic between $x$ and $x'$
(see for instance \cite{dew60,chr76} or the Appendix B of \cite{dor96} for a
review on bitensors).

Using (\ref{eq:xm}) and (\ref{eq:xm'}) we can write $\int _{{t}
'}^{{t}}W({\zeta})d{\zeta}^*$ and
$\left[W({t} )W({t} ')\right]^{-1/2}$,
in (\ref{eq:G(1)2}), in powers of
$\epsilon$. Note also that 
$x^i-{x'}^{i}=2\,\epsilon\, x^i_1 +({\epsilon ^3/ 3})\, 
x^i_3+({\epsilon ^5/ 60})\,
x^i_5+O(\epsilon ^7)$, where we denote $x^1=x,\; x^2=y,\; x^3={z}$. 
With these expressions we can expand the exponential term 
in (\ref{eq:G(1)2}) in powers of $\epsilon$ as
 
$$ {\rm e}^{-i\int _{{t} '}^{{t}}W({\zeta})d{\zeta}^*+ik_x (x-x')+ik_y (y-y')+
ik_3 ({z} -{z} ')}=
{\rm e}^{-i\epsilon\left(\delta _0\,\omega-\delta _1\, k_x-
\delta _2\, k_y-\delta _3\, k_{z}\right)}\,\times\, 
\left(
1+\cdots\right),$$
where we do not expand the
zero adiabatic terms because they will be useful as new integration variables. 
These terms have been included
in the coefficients
$\delta _0$ and $\delta _i$ as,
 
\bb  \delta _0= 2{t} _1^*+{\epsilon ^2\over 3}{t} _3^*+{\epsilon
^4\over 60}{t} _5^*,
\;\;\;\; \delta _i=
2x^i_1+{\epsilon ^2\over 3}x^i_3+{\epsilon ^4\over 60}x^i_5 .        
\label{eq:delta12}\ee 
 
Since the number of derivatives $d/d{t} ^*$ determine the adiabatic order, we 
introduce a new
parameter $T$ which will indicate the adiabatic order. Then at the end
of the calculation we will take $T=1$. With abuse of notation, we
also denote,
$\omega\equiv \omega ({\bar{t}})$,
$W\equiv W({\bar{t}})$.
Recall that in (\ref{eq:Wxi}) $W$ is given
up to adiabatic order four, thus we
have now,
 
$$W=\omega +T^2\,\left({A_2\over\omega ^3}+{B_2\over\omega ^5}\right)+
T^4\,\left({A_4\over\omega ^5}+{B_4\over\omega ^7}+{C_4\over\omega 
^9}+{D_4\over\omega ^{11}}\right),    $$
where the coefficients $\omega$, $A_n$, $B_n$, $C_n$, are given in 
(\ref{eq:heq}) and (\ref{eq:An}).
 
Now, if we substitute these expansions in (\ref{eq:G(1)2}) and
separate the different adiabatic terms by expanding
in powers of the parameter $T$, we can easily see that $G^{(1)}(x,x')$
can be written as a sum of the following type of integrals

\bb
\int ^{\infty}_{-\infty}\int ^{\infty}_{-\infty}\int
^{\infty}_{-\infty}
{dk_x\, dk_y\, dk_{z}}\, {\left[V^{(i)}({\bar{t}})\right]^\alpha\,
\left[V^{(j)}({\bar{t}})\right]^\beta\over\omega ^n}
{\rm e}^{-i\epsilon\left(\delta _0\,\omega-\delta _1\, k_x-
\delta _2\, k_y-\delta _3\, k_{z}\right)},
\label{eq:IntV0}\ee
where,
$V({\bar{t}})=
f_0({\bar{t}})^2+ k_x^2\, f_1({\bar{t}})^2+ k_y^2\, f_2({\bar{t}})^2+
k_{z}^2\, f_3({\bar{t}})^2$. For computational purposes,
these integrals may be solved as
follows. First we use the following change of coordinates,

$$
z_1=f_1\,  k_x,\;\; z_2=f_2\,  k_y,
\;\; z_3=f_3\,  k_z,
\;\;\;\; 
dk_xdk_ydk_{z}=(f_1f_2f_3)^{-1} dz_1dz_2dz_3, 
$$
with $f_i=f_i({\bar{t}})$ and the following definitions,

$$
\Delta _0\equiv \delta _0,\;\;\;
\Delta _i\equiv \delta _i\, f_i^{-1},\;\;\;
\Omega ^2 =f_0^2+z_1^2+z_2^2+z_3^2.$$
We can then use spherical coordinates, $z_1=r\sin\theta\cos\phi$,
$z_2=r\sin\theta\sin\phi$, $z_3=r\cos\theta$, and the following
simplifying notation $\Omega _1\equiv\sin\theta\cos\phi$,
$\Omega _2\equiv\sin\theta\sin\phi$, $\Omega
_3\equiv\cos\theta$. Thus,
$\Omega ^2=f_0^2+r^2$ and we conveniently define,

$$
\Delta\equiv \Delta _1\Omega _1 +\Delta _2\Omega _2 +\Delta _3\Omega _3.
$$
Observe that we can also eliminate the factor $r^2$, which appears
in (\ref{eq:IntV0})
after the change to spherical coordinates, 
using the differential operator $(i\epsilon )^{-1}\partial /\partial\Delta$.
We may finally turn the terms $V^{(i)}({\bar{t}})$, which appear in
(\ref{eq:IntV0}), into the differential operators,

$$
{V}^{(i)}({\bar{t}})\longrightarrow
{\hat V}^{(i)}({\bar{t}})=\left[f_0 ({\bar{t}})^2\right]^{(i)}
+(i\epsilon)^{-2}
\,
\left(\sum _k {\left[f_k ({\bar{t}})^2\right]^{(i)}\over f_k ({\bar{t}})^2}
\,\Omega _k^2\right)\, {\partial ^2\over\partial\Delta ^2}.
$$
After all these changes, the integral (\ref{eq:IntV0}) transforms into

\bb
(i\epsilon )^{-2}\,
\left[ f_1({\bar{t}})f_2({\bar{t}})f_3({\bar{t}}) \right]^{-1}
\int ^{2\pi}_{0}d\phi\int ^{\pi /2}_{0} d\theta\sin\theta\,
\left[{\hat V}^{(i)}({\bar{t}})\right]^\alpha\,\left[{\hat
V}^{(j)}({\bar{t}})\right]^\beta\,
{\partial\over\partial\Delta ^2}{\cal I}_n,
\label{eq:IntV}\ee
where for simplicity we have conveniently defined,

$$
{\cal I}_n=\int ^{\infty}_{-\infty}
{dr\over\Omega ^n}
{\rm e}^{-i\epsilon\left(\Delta _0\,\Omega-\Delta\, r\right)},
$$
and the solution of these ${\cal I}_n$ integrals are discussed in Appendix B.

Solving integrals (\ref{eq:IntV}) and then using (\ref{eq:sigeps}),
(\ref{eq:delta12}), the relation  between the geodesic coefficients
${t} _n^*$ and $x^i_n$ in terms of ${t} ^*_1$ and $x^i_1$ (see
Appendix C) and after performing a rather tedious calculation, the Hadamard
function reduces to,
 
\bba G^{(1)}(x,x')&=&{\bar A}+\sigma\,{\bar b}
+{C}_{{\bar \alpha}{\bar \beta}}\,{\sigma}^{\bar \alpha}{\sigma}^{\bar\beta}
+{D}_{{\bar \alpha}{\bar \beta}{\bar \gamma}{\bar \delta}}\,
{\sigma}^{\bar \alpha}{\sigma}^{\bar\beta}{\sigma}^{\bar
\gamma}{\sigma}^{\bar\delta}
+{1\over 8\pi ^2}\left\{
-{2\over\sigma}
-2{\Delta ^{(2)}}_{{\bar\mu}{\bar\nu}}\,{\sigma ^{\bar\mu}\sigma
^{\bar\nu}\over\sigma}
\right.
\label{eq:Hada2}\\
& & \left.
-2{\Delta ^{(4)}}_{{\bar\mu}{\bar\nu}{\bar\rho}{\bar\tau}}
\,{\sigma ^{\bar\mu}\sigma ^{\bar\nu}\sigma ^{\bar\rho}\sigma
^{\bar\tau}\over\sigma}
-a_1^{(0)}\, L
-\left[
\left(
a_1^{(0)}{\Delta ^{(2)}}_{{\bar\mu}{\bar\nu}}
+{a_1 ^{(2)}}_{{\bar\mu}{\bar\nu}}
\right)\sigma ^{\bar\mu}\sigma ^{\bar\nu}
-{1\over 2}a_2^{(0)}\sigma
\right]\, L
\right\}
\nonumber\eea
where $L$ is a logarithmic term defined as 
$L=2\gamma+\ln (\sigma\,\xi R/2)$,
being $\gamma$ Euler's constant and  where all the involved
coefficients
$\bar A$, $\bar b$, ${C}_{{\bar \alpha}{\bar \beta}}$... are given in Appendix A.
 
According to (\ref{eq:GB}), the Hadamard function can be regularized using the 
elementary
Hadamard solution (\ref{eq:Hada1}) and finally the regularized expression for
$G^{(1)}(x,x')$ up to order $\epsilon ^2$ is,
 
\bba G^{(1)}(x,x')&=&{\bar A}+\sigma\,{\bar B}
+{C}_{{\bar \alpha}{\bar \beta}}\,{\sigma}^{\bar \alpha}{\sigma}^{\bar \beta}
+{D}_{{\bar \alpha}{\bar \beta}{\bar \gamma}{\bar \delta}}\,
{\sigma}^{\bar \alpha}{\sigma}^{\bar\beta}{\sigma}^{\bar
\gamma}{\sigma}^{\bar\delta}
\nonumber\\
& &+{1\over 8\pi ^2}\left\{
-a_1^{(0)}\, {\hat L}
-\left[
\left(
a_1^{(0)}{\Delta ^{(2)}}_{{\bar\mu}{\bar\nu}}
+{a_1 ^{(2)}}_{{\bar\mu}{\bar\nu}}
\right)\sigma ^{\bar\mu}\sigma ^{\bar\nu}
-{1\over 2}a_2^{(0)}\sigma
\right]\, {\hat L}
\right\},
\label{eq:G(1)fR}\eea
where ${\hat L}$ is a bounded logarithmic term given by,
${\hat L}=2\gamma+\ln (\mu ^{2}\,\xi R/2)$, $\mu$ being the arbitrary 
length parameter introduced in (\ref{eq:Hada1}),
and where the coefficient $\bar B$ is 
${\bar B}=b+3\, a_2^{(0)}/(32\pi ^2)$, which is given also in Appendix A.
From (\ref{eq:G(1)fR}) we can directly read off the regularized mean
square field in  
the ``in" vacuum state as
$\langle\phi ^2\rangle ={\bar A}/2-a_1^{(0)}{\hat L}/(16\pi ^2)$.
It is important to remark that the term 
${D}_{{\bar \alpha}{\bar \beta}{\bar \gamma}{\bar \delta}}$ in
(\ref{eq:G(1)fR}) 
appears only as a consequence of our approximate procedure of
calculating the Hadamard function, i.e.  using an adiabatic order four
expansion for the initial modes in powers of the 
mode frequency $\omega ({t})$ and its derivatives.
Had we used an exact expression for the
initial modes (or an adiabatic vacuum state \cite{bir82}), such
a term would  not appear.

\section{Expectation value of the stress-energy tensor}

To calculate the vacuum expectation value of the stress-energy
tensor we have to apply the differential operator 
(\ref{eq:Dopdif}) to
(\ref{eq:G(1)fR}). As we have already pointed out, this is not straightforward 
because we
work with nonlocal quantities. Note first that the operator (\ref{eq:Dopdif}) 
acts on bitensors which
depend on the end points
$x$ and $x'$, but the expression (\ref{eq:G(1)fR}) for $G^{(1)}_B$ depends on 
the midpoint $\bar x$. This means
that we need to covariantly expand (\ref{eq:G(1)fR}) in terms of 
the endpoints $x$ and $x'$.
Consider the general expression, 
 
\bb G^{(1)}_B(x,x')={\bar A}+\sigma\,{\bar B}+
{C}_{{\bar\alpha}{\bar\beta}}\,{\sigma}^{\bar\alpha}{\sigma}^{\bar\beta}
+{D}_{{\bar \alpha}{\bar \beta}{\bar \gamma}{\bar \delta}}\,
{\sigma}^{\bar \alpha}{\sigma}^{\bar\beta}{\sigma}^{\bar
\gamma}{\sigma}^{\bar\delta}
+{\hat L}\,{E}_{{\bar \alpha}{\bar \beta}}\,{\sigma}^{\bar
\alpha}{\sigma}^{\bar 
\beta}
,\label{eq:GABCD}\ee
where it is understood that ${\bar A}$, ${\bar B}$, 
${C}_{{\bar\alpha}{\bar\beta}}$, 
$D_{{\bar\alpha}{\bar\beta}{\bar\gamma}{\bar\delta}}$,
$E_{{\bar\alpha}{\bar\beta}}$ are functions 
that depend on the endpoints $x$, $x'$ but are evaluated at the midpoint $\bar 
x$. Then, following the
formalism of Appendix B of reference \cite{dor96}, we can expand in
the neighbourhood of $x$ as, 
 
\bb G^{(1)}_B(x,x')={\bar A}+\sigma\,{\bar B}+
{{\bar g}^{\bar\alpha}}_{\alpha}\,{{\bar g}^{\bar\beta}}_{\beta}\,
{C}_{{\bar\alpha}{\bar\beta}}\,{\sigma}^{\alpha}{\sigma}^{\beta}
+{{\bar g}^{\bar\alpha}}_{\alpha}\,{{\bar g}^{\bar\beta}}_{\beta}\,
{{\bar g}^{\bar\gamma}}_{\gamma}\,{{\bar g}^{\bar\delta}}_{\delta}
{D}_{{\bar \alpha}{\bar \beta}{\bar \gamma}{\bar \delta}}\,
{\sigma}^{\alpha}{\sigma}^{\beta}{\sigma}^{
\gamma}{\sigma}^{\delta}
+{{\bar g}^{\bar\alpha}}_{\alpha}\,{{\bar g}^{\bar\beta}}_{\beta}\,
{\hat L}\,{E}_{{\bar \alpha}{\bar \beta}}\,{\sigma}^{\alpha}{\sigma}^{
\beta}
 ,\label{eq:GABCD1}\ee  
where before expanding, a homogeneization of the indices by the parallel 
transport bivector
${{\bar g}^{\mu}}_{\bar\nu}$, i.e, $\sigma ^{\mu}={{\bar 
g}^{\mu}}_{\bar\nu}\,\sigma ^{\bar\nu},$
has been applied. The covariant expansions at $x$ are given by,
 
\bba {\bar A} &=& A-{1\over 2}\, A_{;\alpha}\,{\sigma}^{\alpha}+
{1\over 8}\, A_{;\alpha\beta}\,{\sigma}^{\alpha}\,{\sigma}^{\beta}+O(\epsilon 
^3),
\;\;\;\; {\bar B}=B+O(\epsilon ),
\label{eq:CovAB}\\
{{\bar g}^{\bar\alpha}}_{\alpha}\,{{\bar 
g}^{\bar\beta}}_{\beta}\,{C}_{{\bar\alpha}{\bar\beta}} &=&
{C}_{{\alpha}{\beta}}+O(\epsilon ),\;\;\;\;
{{\bar g}^{\bar\alpha}}_{\alpha}\,{{\bar g}^{\bar\beta}}_{\beta}\,
{{\bar g}^{\bar\gamma}}_{\gamma}\,{{\bar g}^{\bar\delta}}_{\delta}
{D}_{{\bar \alpha}{\bar \beta}{\bar \gamma}{\bar \delta}}
=D_{{\alpha}{\beta}{\gamma}{\delta}}+O(\epsilon )
\;\;\;\;
{{\bar g}^{\bar\alpha}}_{\alpha}\,{{\bar g}^{\bar\beta}}_{\beta}\,
E_{{\bar\alpha}{\bar\beta}}=
E_{{\alpha}{\beta}}+O(\epsilon ).
\nonumber\eea
 
Now we can apply the differential operator (\ref{eq:Dopdif}) to 
(\ref{eq:GABCD1}) and we  will consider a {\it conformal coupling}
($\xi =1/6$), 
which, from general grounds,  should provide a good qualitative
description for photons. 
If we introduce the
operators,
 
\bb {\cal L}^{(1)}_{\mu\nu}=\nabla _{\mu '}\nabla _{\nu}+\nabla _{\nu '}\nabla 
_{\mu },
\;\;\;\; {\cal L}^{(2)}_{\mu\nu}=\nabla _{\mu '}\nabla _{\nu '}+\nabla _{\nu 
}\nabla _{\mu },
\label{eq:L12opdif}\ee
the operator (\ref{eq:Dopdif}), for the conformal case, is
 
$${\cal D}^{\xi =1/6}_{\mu\nu}={1\over 12}\,\left[2\,\left({\cal 
L}^{(1)}_{\mu\nu}-
{1\over 4}\, g_{\mu\nu}\, g^{\alpha\beta}\,{\cal L}^{(1)}_{\alpha\beta}\right)
-\left({\cal L}^{(2)}_{\mu\nu}-
{1\over 4}\, g_{\mu\nu}\, g^{\alpha\beta}\,{\cal L}^{(2)}_{\alpha\beta}\right)
\right]
-{1\over 12}\,\left(R_{\mu\nu}-{1\over 4}\, R\, g_{\mu\nu}\right)
.$$
By the properties (\ref{eq:sigeps}) of the geodetic interval bivector
$\sigma ^{\mu}$ we can prove the following identities,

$$\lim _{x\rightarrow x'}{\cal L}^{(1)}_{\mu\nu}\,\sigma =
-\lim _{x\rightarrow x'}{\cal L}^{(2)}_{\mu\nu}\,\sigma =-2\, 
g_{\mu\nu},\;\;\;\; \lim _{x\rightarrow x'}{\cal
L}^{(1)}_{\mu\nu}\,{\sigma}^{\alpha}{\sigma}^{\beta}= -\lim _{x\rightarrow 
x'}{\cal
L}^{(2)}_{\mu\nu}\,{\sigma}^{\alpha}{\sigma}^{\beta}= 
-4\,\delta^{\alpha}_{(\mu}\,\delta^{\beta}_{\nu )},$$

\bban \lim _{x\rightarrow x'}{\cal L}^{(1)}_{\mu\nu}\,
{{\sigma}^{\alpha}{\sigma}^{\beta}{\sigma}^{\gamma}{\sigma}^{\delta}\over%
\sigma}&=&
-\lim _{x\rightarrow x'}{\cal L}^{(2)}_{\mu\nu}\,
{{\sigma}^{\alpha}{\sigma}^{\beta}{\sigma}^{\gamma}{\sigma}^{\delta}\over%
\sigma}=
\\
&&
-{48\over\Sigma}\,\delta^{(\alpha}_{(\mu}\,\delta^{\beta}_{\nu )}\, 
p^{\gamma}p^{\delta )}
+64\,\delta^{(\alpha}_{(\mu}\, p^{\beta}_{\nu )}\, p^{\gamma}p^{\delta )}
+8\, 
p^{\alpha}p^{\beta}p^{\gamma}p^{\delta}\,\left[g_{\mu\nu}-4{p_{\mu}p_{\nu}\over%
\Sigma}\right],\eean
where $(\cdots)$ is the usual symmetrization operator. Note that by a 
straightforward application of Synge's
theorem (see for example references \cite{dew60,chr76} or Appendix B
of reference \cite{dor96}), the coincidence limits of 
${\cal L}^{(1)}_{\mu\nu}$ and ${\cal L}^{(2)}_{\mu\nu}$ differ in a sign when 
they are applied on bitensors
for which the first covariant derivative has null coincidence limit. But when 
they are applied
on the biscalar $\bar A$ in (\ref{eq:CovAB}) they coincide:
$\lim _{x\rightarrow x'}{\cal L}^{(1)}_{\mu\nu}\,{\bar A}=
\lim _{x\rightarrow x'}{\cal L}^{(1)}_{\mu\nu}\,{\bar A}=A_{;\mu\nu}/2$.
The application of the opperators (\ref{eq:L12opdif})
over quartic products of $\sigma ^{\mu}$ gives  path dependent terms when the 
limit $x\rightarrow x'$
is performed. Therefore an appropriate averaging  
procedure is required. The most elementary averaging is called 
{\it four dimensional
hyperspherical averaging} \cite{adl77}  and 
consists in giving the same weight to
all the geodesics which emanate from $x$ as follows. First one analytically
continues to a Euclidean metric the components of the tangent vectors to the 
geodesics which emanate from
$x$. Second, one averages over a 4-sphere, and third, 
the results are continued 
back to the original
metric. It is not very complicated to find the following averaging formulae,

$$ \langle {\sigma}_{\mu}{\sigma}_{\nu} \rangle ={1\over 
2}\,\sigma\,g_{\mu\nu},\;\;\;\; \langle
{\sigma}_{\mu}{\sigma}_{\nu} {\sigma}_{\tau}{\sigma}_{\rho} \rangle = {1\over 
2}\,\sigma ^2\,
g_{\mu (\nu} g_{\tau\rho )}, \;\;\;\;
 \langle {\sigma}_{\mu}{\sigma}_{\nu}
{\sigma}_{\tau}{\sigma}_{\rho}{\sigma}_{\xi}{\sigma}_{\eta}
\rangle = {5\over 8}\,\sigma ^3\,  
g_{\mu (\nu}g_{\tau (\rho}g_{\xi\eta ))}.
$$
Note that from the symmetry of the averaging procedure, the average of an odd 
number of components $\sigma
^{\mu}$ vanishes identically.
 
We now have all we need to evaluate the vacuum expectation value of the 
stress-energy tensor
by application of the differential operator (\ref{eq:Dopdif}) to the 
expression (\ref{eq:GABCD1}) of
$G^{(1)}_B(x,x')$. Observe for example that 
${\cal D}^{\xi =1/6}_{\mu\nu}\sigma =0$.
Then, in
the orthonormal basis 
$\theta _1={g^{1/2}_{{t}{t}}}\,d{t}$,
$\theta _2={g^{1/2}_{{z}{z}}}\, d{z}$, $\theta _3={g^{1/2}_{xx}}\, dx$, 
$\theta _4={g^{1/2}_{yy}}\, dy$,  using the trace anomaly
prescription (\ref{eq:GBT}),  we obtain
the following  expectation values $\langle T_{\mu\nu}\rangle$ in
the conformal coupling case and for values
$0<\epsilon<{t}<\pi /2$ of coordinate ${t}$
 
\bba\langle T_{\mu\nu}\rangle &=&
{2\gamma+\ln (\mu ^{2}\,\xi R/2)\over 2880\, (L_1L_2)^2\, 
(1+z)^4}\,{\rm diag}(-1,\, -1,\, 1,\, -1)
-{1\over 4}\langle T^{\tau}_{\tau}\rangle g_{\mu\nu}+
\nonumber
\\
&&
{\rm 
diag}\left(\rho _1({t}) ,\; -\rho _2({t}) ,\; \rho _1({t})+2\rho
_2({t}),\; -\rho _2({t})\right),
\label{eq:TMNconforme}\eea
where we have used for simplicity the
notation $z=\sin{t}$. 
The trace anomaly term in that case is given
by  (\ref{eq:GBT}) as,

$${1\over 4}\langle T^{\tau}_{\tau}\rangle g_{\mu\nu} =
{a_2\over 64\,\pi ^2}\, g_{\mu\nu}=
{23-2\, z-z^2\over 2880\, (L_1L_2)^2\, \pi ^2\, (1+z)^6}
{\rm diag}\left(1 ,\, -1 ,\, -1,\, -1\right),$$ 
and the functions $\rho _1({t})$ and $\rho _2({t})$
are  given by,
 
\bb\rho _1({t}) ={30254-106858\, z+168510\, z^2-147520\, z^3+62495\, z^4\over
241920\, (L_1L_2)^2\,\pi ^2\,(1-z)^2(1+z)^6},\label{eq:rho1}\ee
\bb\rho _2({t}) ={-566190+1821200\, z-2415234\, z^2+1710444\,
z^3-472795\, z^4\over 
2661120\, (L_1L_2)^2\, \pi ^2\,(1-z)^2(1+z)^6}.\label{eq:rho2}\ee
Recall that $\rho _1({t})$ is a positive definite function in the
interval $0<\epsilon <{t}<\pi /2$. Both functions are unbounded at the horizon
$({t}=\pi /2)$, and the expectation value of the stress-energy tensor
near the horizon is approximately given by,

\bb\left.\langle T_{\mu\nu}\rangle\right|_{{t}\simeq\pi /2} ={\rm 
diag}\left(\rho ({t}) ,\; -\rho ({t}) ,\; 3\rho ({t})
,\; -\rho ({t})\right)-
{1\over 4}\langle T^{\tau}_{\tau}\rangle g_{\mu\nu},
\label{eq:TMapprox}\ee
where $\rho ({t})=\Lambda (L_1L_2)^{-2}\cos{t}^{-4}$ and $\Lambda\simeq
0.00018$. This result is compatible with our previous result
\cite{dor96}. In that case, however, we found a bigger value for the
parameter $\Lambda\simeq 0.001$. Such a difference makes sense if we
recall that in \cite{dor96} we obtained an approximate value for the
expectation value of the stress-energy tensor, essentially using the
``blueshift'' effect as the initial modes approach the folding
singularities of the interaction region. 
Therefore, such a result was
a good approximation in the region near both the folding singularities
and the Killing-Cauchy horizon,
where a priori we would expect a stronger effect than in the center of the 
collision,  which is the only region considered in the present paper.

For values of $t\leq 0$, $\langle T_{\mu\nu}\rangle =0$, 
and to be consistent with the approximation 
we have used for the space-time geometry, we should
require that the value of $\langle T_{\mu\nu}\rangle$
(\ref{eq:TMNconforme}), which is valid for $0<\epsilon <{t}<\pi /2$, goes 
smoothly to zero as ${t}\rightarrow 0$. In fact, this can be achieved
using an adequate matching of the line element (\ref{eq:dshatIb}) with the
flat line element through the interval $0\leq{t}\leq\epsilon$. Observe that
the coefficient of
the logarithmic term in the stress-energy tensor (\ref{eq:TMNconforme})
deppend only on locally constructed curvature terms (as can be seen
from the Hadamard function  (\ref{eq:G(1)fR})). Therefore,
with an adequate matching of the space-time
geometry, this coefficient will also
smoothly vanish towards the flat space region, below $t=0$.
The details of such a matching, however, will not affect
the main features of the stress-energy tensor (\ref{eq:TMNconforme}),
particularly when the Killing-Cauchy horizon is approached.

We must recall, however, that
although the value (\ref{eq:TMNconforme}) for 
$\langle T_{\mu\nu}\rangle$ satisfies asymptotically 
the conservation equation close to
the Killing-Cauchy horizon, it does not satisfy exactly the conservation
equation throughout region $\cal S$, essentially
because it is obtained by means of an
approximation in the field modes. Nevertheless, we could obtain a
truly conserved $\langle T_{\mu\nu}\rangle$, in the context of the
present approximation, by solving the conservation equation
considering a $\langle {T^{\mu}}_{\nu}\rangle$ given by
$(T^t_t({t}),\,T^z_z({t}),\,T^x_x({t}),\,T^y_y({t}))$ with the
conditions: i) $T^y_y({t})=T^z_z({t})$, which not only is compatible with
(\ref{eq:TMNconforme}) but also holds for the exact
interaction region geometry, ii) trace anomaly condition, i.e. 
$T^x_x({t})=\langle T^{\mu}_{\mu}\rangle
-T^t_t({t})-2\, T^z_z({t})$, iii) the ansatz $T^t_t({t})=\rho
_1({t})$, which is the approximate value of $T^t_t({t})$ obtained in
our calculation. Finally, the conservation equation gives
straightforwardly a value for $T^z_z({t})$ and $T^x_x({t})$
which are compatible with the value $T^z_z({t})=\rho _2({t})$ and
$T^x_x({t})=-\rho _1({t})-2\rho _2({t})$ obtained in our
approximation. In particular, they have the same behaviour near the
Killing-Cauchy horizon.
 
Inspection of (\ref{eq:TMNconforme}) shows that not only is 
the {\em weak energy
condition} satisfied \cite{wal84}, which means that the energy density is 
nonnegative for any observer, but also
the {\em strong energy condition} is satisfied.

\section{Conclusions}
 
We have calculated the expectation value of the stress-energy tensor
of a massless scalar field in a space-time representing the head on
collision of two gravitational plane waves throughout the causal
past of the collision center and in the field state which
corresponds to the physical vacuum state before the collision takes
place. We have performed the calculations in this particular region
essentially because we could introduce a suitable approximation to the
space-time metric (see Fig. 3) which not only allowed us to
dramatically simplify the calculations but also to keep unchanged the
main physical features, in particular the behaviour of the
stress-energy tensor near the Killing-Cauchy horizon of the
interaction region. In fact, such an approximation is also valid for more
generic plane wave space-times, and this will be the subject of a
forthcoming paper.

The results we have obtained are the following: before the collision
of the waves $\langle T_{\mu\nu}\rangle =0$, which correspond to the
lower edge Fig. 3. Then, after the collision the value of 
$\langle T_{\mu\nu}\rangle$ starts to increase until it grows
unbounded towards the Killing-Cauchy horizon of the interaction
region. The 
weak energy condition is satisfied, the rest energy density is positive and 
diverges as $\cos ^{-4}{t}$. Two of 
the principal pressures are negative and of the same order of magnitude 
of the energy density. The
strong energy condition is also satisfied, 
$\langle T^{\mu}_{\mu}\rangle$ is finite but  $\langle
T^{\mu\nu}\rangle \langle T_{\mu\nu}\rangle$ diverges at the horizon and we 
may
use ref \cite{hel93} on the stability of Cauchy horizons to argue that the 
horizon will aquire by backreaction
a curvature singularity.
Thus, contrary to  simple plane waves,
which do not polarize the vacuum \cite{des75,gib75}, the nonlinear collision 
of these waves polarize the vacuum
and the focusing effect that the waves exert produce at the focusing
points an unbounded positive energy density.
Therefore, when the colliding waves produce a Killing-Cauchy horizon, that 
horizon is unstable by vacuum
polarization.
 
In the more generic case when the wave collision produces a spacelike 
singularity it seems clear that the
vacuum expectation value of the stress-energy tensor will also grow unbounded 
near the singularity. In fact, in a forthcoming paper we will extend
the approximation introduced in the present work to a more generic
plane wave spacetime with the objective of more generally proving that the
negative pressures associated to the quantum fields could not prevent
the formation of the singularity.

\vskip 1.25 truecm
 
{\Large{\bf Acknowledgements}}
 
\vskip 0.5 truecm
 
\noindent
I am  grateful to R. M. Wald, R. Geroch, E. Verdaguer, A. Campos,
E. Calzetta, A. Feinstein,  
J. Iba{\~n}ez and A. Van Tonder for helpful
discussions. This work has been partially supported by 
NSF grant PHY 95-14726 to The University of Chicago and by the Direcci\'o
General de Recerca de la
Generalitat de Catalunya through the grant 1995BEAI300165.

\appendix

\section{Useful  adiabatic expansions}

The coefficients for the Hadamard function in (\ref{eq:G(1)fR}), using
for simplicity the notation $z=\sin{t}$, are:
 
\bban {\bar
A}&=&{6+z\over 60\, L_1L_2\, \pi ^2\, (1+z)^3},
\\
& &
\\
{\bar B}&=&
{1859-9652\, z+17478\, z^2-14644\, z^3+5543\, z^4\over
60480\, (L_1L_2)^2 \, \pi ^2\, (1-z)^2(1+z)^6}
\\
& &
\\
C_{{\bar z}{\bar z}}&=&
{-20959+1075\, z+122490\, z^2-174964\, z^3+76711\, z^4\over
166320\, L_1L_2\, \pi ^2\, (1-z)^2(1+z)^4}
\\
& &
\\
C_{{\bar x}{\bar x}}&=&
{30629-127830\, z+229803\, z^2-212506\, z^3+88518\, z^4\over
332640\,  (L_1L_2)^2\, \pi ^2\, (1-z)(1+z)^7}
\\
& &
\\
C_{{\bar y}{\bar y}}&=&{1\over L_1L_2}\, C_{{\bar z}{\bar z}}
\\
& &
\\
D_{{\bar z}{\bar z}{\bar z}{\bar z}}&=&
{282-1039\, z+1656\, z^2-1407\, z^3+516\, z^4
\over 3840\, \pi ^2\, (1-z)^2(1+z)^2}
\\
& &
\\
D_{{\bar x}{\bar x}{\bar x}{\bar x}}&=&
{1+10\, z-14\, z^2-6\, z^3+39\, z^4
\over 960\, (L_1L_2)^2\, \pi ^2\, (1+z)^8}
\\
& &
\\
D_{{\bar y}{\bar y}{\bar y}{\bar y}}&=&
{1\over (L_1L_2)^2}\, D_{{\bar z}{\bar z}{\bar z}{\bar z}}
\\
& &
\\
D_{{\bar z}{\bar z}{\bar x}{\bar x}}&=&
{-83+108\, z+190\, z^2-426\, z^3+237\, z^4
\over 5760\, L_1L_2\, \pi ^2\, (1-z)(1+z)^5}
\\
& &
\\
D_{{\bar z}{\bar z}{\bar y}{\bar y}}&=&
{-184-183\, z+1780\, z^2-2359\, z^3+954\, z^4
\over 11520\, L_1L_2\, \pi ^2\, (1-z)^2(1+z)^2}
\\
& &
\\
D_{{\bar x}{\bar x}{\bar y}{\bar y}}&=&
{1\over L_1L_2}\, D_{{\bar z}{\bar z}{\bar x}{\bar x}}
\eean

The coefficients for the {\em midpoint
expansion} of the locally constructed Hadamard function (\ref{eq:Hada1})
are:

\bban
a_1^{(0)}&=&-R\,\left(\xi-{1\over 6}\right),\;\;\;\;
{\Delta ^{(2)}}_{\mu\nu}={1\over 12}R_{\mu\nu},
\\
& &
\\
a_2^{(0)}&=&
{1\over 2}\left({1\over 6}-\xi\right)^2\, R^2
+{1\over 6}\left({1\over 5}-\xi\right)\, {R_{;\alpha}}^\alpha
-{1\over 180}\, R^{\alpha\beta} R_{\alpha\beta}
+{1\over180}\,R^{\alpha\beta\gamma\delta}R_{\alpha\beta\gamma\delta},
\\
& &
\\
{a_1^{(2)}}_{\mu\nu}&=&
{1\over 24}\left({1\over 10}-\xi\right)\, R_{;\mu\nu}
+{1\over 120}\, {R_{\mu\nu ;\alpha}}^\alpha
-{1\over 90}\, {R^\alpha}_\mu R_{\alpha\nu}+
\\
& &
\\
& & {1\over 180}\, R^{\alpha\beta}R_{\alpha\mu\beta\nu}
+{1\over 180}\, {R^{\alpha\beta\gamma}}_{\mu}\,  
R_{\alpha\beta\gamma\nu},
\\
& &
\\
{\Delta ^{(4)}}_{\mu\nu\rho\tau}&=&
 {3\over 160}\, R_{\mu\nu ;\rho\tau}
+{1\over 288}\, R_{\mu\nu}R_{\rho\tau}
+{1\over 360}\, {{{R^\alpha}_\mu}^\beta}_\nu  
R_{\alpha\rho\beta\tau}.
\eean

\section{Some useful integrals.}
 
Let us define
\bb  {\cal I}_{n}=\int _{-\infty}^{\infty} {dk}\,\Omega ^n\,
{\rm e}^{-i\epsilon\left({\Delta} _0\,\Omega -{\Delta} _1\, k\right)}, \;\;\;\;
\Omega =\sqrt{k^2+\lambda};\;\;\; \lambda >0. \label{eq:In}\ee
For the particular value $n=-1$ this integral can be easily solved. Performing 
the following change of
variable,
 
\bb t=t(k)={\Delta} _0\,\Omega-{\Delta}_1\, k={\Delta} 
_0\,\sqrt{k^2+\lambda}-{\Delta} _1\, k,    \label{eq:t(k)}\ee
we must consider separately the two possibilities $|{\Delta} _0|>|{\Delta} _1|$ or 
$|{\Delta} _0|<|{\Delta} _1|$.
For $|{\Delta} _0|>|{\Delta} _1|$ and
without loss of generality we may take
${\Delta} _0,\;{\Delta} _1>0$, and invert the change 
(\ref{eq:t(k)}) as,
 
\bb  k=k(t)=\gamma\,\left[{\Delta} _1\, t\pm{\Delta} _0\,\sqrt{t^2-\beta 
^2}\right],   \label{eq:k(t)}\ee
where we use two new variables
$\gamma$ and $\beta$, defined as,
 
\bb \gamma ^{-1}\equiv {\Delta} _0^2-{\Delta} _1^2,\;\;\; \beta ^2\equiv{\lambda 
\gamma ^{-1}}.              
\label{eq:gam-bet}\ee
Note that $k(t)$, in (\ref{eq:k(t)}), is a bivalued function and we have to be 
careful
in changing the integration limits. Since $|{\Delta} _0|>|{\Delta} _1|$, 
then $\lim _{k\rightarrow\pm\infty} t(k)=+\infty$,
and this means that (\ref{eq:t(k)}) has an absolute minimum, which is
at the point 
$(+\gamma\,\beta\,{\Delta} _1,\;\beta)$. With this it is easy to see that we 
have to take as inverse function of
(\ref{eq:t(k)}) the function (\ref{eq:k(t)}) with the plus sign to the right 
of the minimum $t=\beta$ and
with the minus sign to the left of the minimum. Therefore we can split the 
integral
(\ref{eq:In}), for $n=-1$, into two parts on each side of $t=\beta$, i.e,
 
\bb{\cal I}_{-1}=\int _{-\infty}^{\gamma\beta{\Delta} _1} {dk\over\Omega}\,
{\rm e}^{-i\epsilon\left({\Delta} _0\,\Omega -{\Delta} _1\, k\right)}+
\int _{\gamma\beta{\Delta} _1}^{\infty} {dk\over\Omega}\,
{\rm e}^{-i\epsilon\left({\Delta} _0\,\Omega -{\Delta} _1\, k\right)}. 
\label{eq:I-11}\ee
The change of variables (\ref{eq:t(k)}) gives
$\Omega ^{-1}{dk}=\pm (t^2-\beta ^2)^{-1/2}dt$,
with the minus sign for the first integral in (\ref{eq:I-11}) and the plus 
sign for the second integral. Then
the integration can be easily performed in terms of a zeroth order Bessel 
function \cite{gra80}, as
 
\bb {\cal I}_{-1}=2\,\int _{\beta}^{\infty} 
{{\rm e}^{-i\epsilon\, t}\over\sqrt{t^2-\beta ^2}}\, dt=
2\,{\rm K}_0 \left(i\epsilon\,\beta\right). \label{eq:I-1R}\ee
 
For the case
$|{\Delta} _0|<|{\Delta} _1|$, we define the parameter $\gamma$, as,
$\gamma ^{-1}\equiv {\Delta} _1^2-{\Delta} _0 ^2$,
and the inverse function of (\ref{eq:t(k)}) is,
 
\bb  k=k(t)=-\gamma\,\left[{\Delta} _1\, t\pm{\Delta} _0\,\sqrt{t^2+\beta 
^2}\right],   \label{eq:k(t)'}\ee
which is again a bivalued function. Now, however, the function
(\ref{eq:t(k)}) has no extrema, and it is easy to see that we can take a 
single inverse everywhere as,
 
$$ k(t)=-\gamma\,\left[{\Delta} _1\, t - {\Delta} _0\,\sqrt{t^2+\beta ^2}\right],
\;\;\;\; {dk\over\Omega}=-{dt\over\sqrt{t^2+\beta ^2}},$$
so that (\ref{eq:In}), for $n=-1$, can also be integrated in terms of a zeroth 
order Bessel function
\cite{gra80} as,
 
\bb {\cal I}_{-1}=\int _{-\infty}^{\infty} {dk\over\Omega}\,
{\rm e}^{-i\epsilon\left({\Delta} _0\,\Omega -{\Delta} _1\, k\right)}=
\int _{-\infty}^{\infty} 
{{\rm e}^{-i\epsilon\, t}\over\sqrt{t^2+\beta ^2}}\, dt=
2\,\int _{0}^{\infty} 
{\cos\left(\epsilon\, t\right)\over\sqrt{t^2+\beta ^2}}\, dt=
2\,{\rm K}_0 \left(\epsilon\,\beta\right).\label{eq:I-1R'}\ee
 
Finally putting together the results
(\ref{eq:I-1R}) and (\ref{eq:I-1R'}), we can write
 
\bb {\cal I}_{-1}=2\,{\rm K}_0 \left(i\epsilon\,\beta\right),  
\label{eq:I-1RF}\ee
with the parameter $\beta$ given by,
\bb \beta ^2={\lambda \gamma ^{-1}}=\lambda\,\left({\Delta} _0^2-{\Delta} 
_1^2\right). \label{eq:gam-betR}\ee
Recall however that we have only considered the case $\lambda >0$. The case
$\lambda <0$ involves more delicate complex contour integrations, but the
final result is the same.
 
The integrals ${\cal I}_{n}$ with $n\leq -2 $
are finite in the limit $\epsilon\rightarrow 0$ and 
they  can also be recursively calculated from ${\cal I}_{-1}$
and the following recursion relation for $n\leq -1$:
 
\bb {\partial{\cal I}_{n}\over\partial\lambda}={n\over 2}\,{\cal I}_{n-2}-
i{\epsilon\,{\Delta} _0\over 2}\,{\cal I}_{n-1}.              
\label{eq:rec-I-n}\ee
This allows us to calculate all the ${\cal I}_n$ with $n\leq -1$ only if we 
know at least
another integral besides ${\cal I}_{-1}$. Fortunately it is not very difficult 
to evaluate
${\cal I}_{-2}$ from the recursion relation
$${\cal I}_{n}={i\over\epsilon}\,{\partial{\cal I}_{n-1}\over\partial{\Delta} 
_0},$$
and by a simple integration,
 
$${\cal I}_{-2}({\Delta} _0)=-i\epsilon\,\int _0^{{\Delta} _0}\, d{\Delta} 
_0'\,{\cal I}_{-1}({\Delta} _0')+
{\cal I}_{-2}(0), $$
where the integration constant
${\cal I}_{-2}(0)$, which corresponds to the value of 
${\cal I}_{-2}$ at the special point ${\Delta} _0=0$,
can be calculated directly from (\ref{eq:In}) \cite{gra80} as
 
$$ \left.{\cal I}_{-2}(0)\right|_{\lambda >0}=
{\pi \over\lambda ^{1/2}}\, {\rm e}^{-\epsilon\,\lambda ^{1/2}\,{\Delta} 
_1},\;\;\;\; 
\left.{\cal I}_{-2}(0)\right|_{\lambda <0}=-
{\pi \over (-\lambda )^{1/2}}\, 
\sin \left[{\epsilon\, (-\lambda )^{1/2}\,{\Delta} 
_1}\right].$$

\section{Geodesic coefficients ${t} _n^*$ and $x_n^i$}

We start with either a timelike or spacelike geodesic connecting the points 
$x'$, $\bar x$ and $x$, which
can be written in terms of the proper geodesic distance $\tau$ as,
 
$$x^{\mu}=x^{\mu}\left({\bar\tau}+\epsilon\right),\;\;\;\;{\bar
x}^{\mu}=x^{\mu}\left({\bar\tau}\right),\;\;\;\;
x'^{\mu}=x^{\mu}\left({\bar\tau}-\epsilon\right),  $$ 
and define at the midpoint
$\bar x$ the parameters
${t} _n^*=\left.{d^n{t} ^* / d\tau ^n}\right|_{\bar\tau}$ and 
$x^i_n=\left.{d^n x^i 
/ d\tau
^n}\right|_{\bar\tau}$, which determine the geodesic.
From our approximate metric (\ref{eq:dsgen}) in the interaction region,
we have

$$
{{t} ^*_1}^2={1\over f_1f_2f_3}+{{z} _1^2\over f_3^2}
+{x_1^2\over f_1^2}
+{y_1^2\over f_2^2},
$$
and since the metric (\ref{eq:dsgen}) does not depend on coordinates
$x$, $y$ or ${z}$, then $p_x=g_{xx}\, x_1$, $p_y=g_{yy}\, y_1$ and
$p_{z}=g_{{z}{z}}\, {z} _1$, are constants. Therefore we can rewrite
the above expression as,

\bb
{{t} ^*_1}^2={1\over f_1f_2f_3}+{p_{z} ^2\over (f_1f_2)^2}
+{p_x^2\over (f_2f_3)^2}
+{p_y^2\over (f_1f_3)^2}\equiv F({t} ^*),
\label{eq:xi*2}
\ee
where for convenience we have introduced the function $F({t} ^*)$.
Then in terms of ${t} ^*_1$ and derivatives of the function $F$, the
${t} ^*_n$ coefficients can be written as,

$$
{t} ^*_2={1\over 2}F',\;\;\;\; {t} ^*_3={1\over 2}{t} ^*_1F'',\;\;\;\;
{t} ^*_4={1\over 4}\left[(FF'')'+FF'''\right],
$$

$$
{t} ^*_5={1\over 4}{t} ^*_1\left[(FF'')'+FF'''\right]',
$$
where we denote $f'=df/d{t}^*$.

Analogously, using that $p_i=g_{ii}\, x^i_1$ are constants, 
the coefficients $x^i_n$ can be easily written in terms of
$x^i_1$, ${t} ^*_1$, $F$ and the metric coefficients as,

$$
x^i_2={{g^{ii}}'\over g^{ii}} {t} ^*_1 x^i_1,\;\;\;\;
x^i_3={1\over 2}{(F{g^{ii}}')'+F{g^{ii}}''\over g^{ii}}x^i_1,\;\;\;\;
x^i_4={1\over 2}{\left[(F{g^{ii}}')'+F{g^{ii}}''\right]'
\over g^{ii}} {t} ^*_1 x^i_1,\;\;\;\;
$$

$$
x^i_5={1\over 4}{x^i_1\over g^{ii}}
\left\{
\left[
F\left(
(F{g^{ii}}')'+F{g^{ii}}''
\right)'
\right]'+
F\left(
(F{g^{ii}}')'+F{g^{ii}}''
\right)''
\right\}.
$$
All these expressions, evaluated at the midpoint $\tau={\bar\tau}$,
produce the geodesic coefficients we need.

\vskip 1.25 truecm
 
{\Large{\bf Figure captions}}
 
\vskip 0.5 truecm
 
\noindent
{\bf Fig. 1}
The colliding wave space-time consists of two
approaching waves, regions II and III, in a flat background, region
IV, and an interaction region, region I. The two waves move in the
direction of two null coordinates $u$ and $v$.
The four space-time regions
are separated by the two null wave fronts $u=0$ and $v=0$. The
boundary between regions I and II is  $\{0\leq u<\pi /2,\; v=0\}$, the
boundary between regions I and III is $\{u=0,\; 0\leq v <\pi /2\}$, and
the boundary of regions II and III with region IV is
$\{u\leq 0,\;v=0\}\cup\{u=0,\;v\leq 0\}$. Region I meets region IV
only at the surface $u=v=0$. The Killing-Cauchy horizon in the region
I corresponds to the hypersurface $u+v=\pi /2$ and plane wave regions
II and III meet such a Killing-Cauchy horizon
only at ${\cal P}=\{u=\pi /2,\; v=0\}$ and
${\cal P}'=\{u=0,\; v=\pi /2\}$ respectively. 
The hypersurfaces  $u=\pi /2$ in region II  and $v=\pi /2$ 
in region III are a type of topological singularities commonly referred as
folding singularities and they
must be identified with $\cal P$ and ${\cal P}'$ respectively.

\vskip 0.5 truecm

\noindent
{\bf Fig. 2}
The subset of Cauchy data which affects the evolution of the 
quantum field along the center $u=v$ of the plane wave  collision lies on
the segments $\{0\leq u<\pi /4,\; v=0\}\cup\{u=0,\; 0\leq v <\pi /4\}$.
Region $\cal S$ is the causal future of this Cauchy data (or
equivalently, the causal past of the collision center).

\vskip 0.5 truecm

\noindent
{\bf Fig. 3}
We change the {\em mode propagation problem} for the plane wave
collision, in the causal past of the collision center, region $\cal S$,
by a much simpler Schr\"odinger-type problem which consists
in eliminating the dependence of the field equation on coordinate
$z$ by taking $z=0$, and substituting the Cauchy data which
come from the single plane wave regions on segments
$\{0\leq u<\pi /4,\; v=0\}\cup\{u=0,\; 0\leq v <\pi /4\}$ 
by much simpler Minkowski Cauchy data. This procedure is essentially
equivalent to modifying the space-time geometry in the causal
past of the collision center by eliminating
the dependence on coordinate $z$ in the
line element, setting $z=0$, and smoothly matching this
line element, through plane wave regions II and III, 
with the flat spacetime below the segment
$\{{t}=0,\; -\pi /4< v < \pi /4\}$.

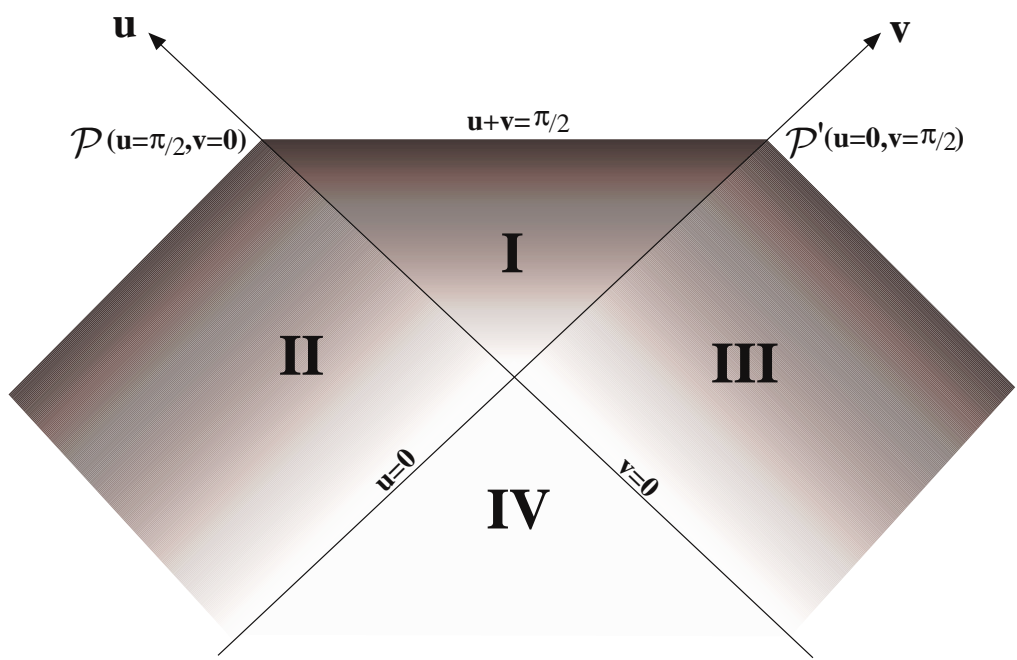
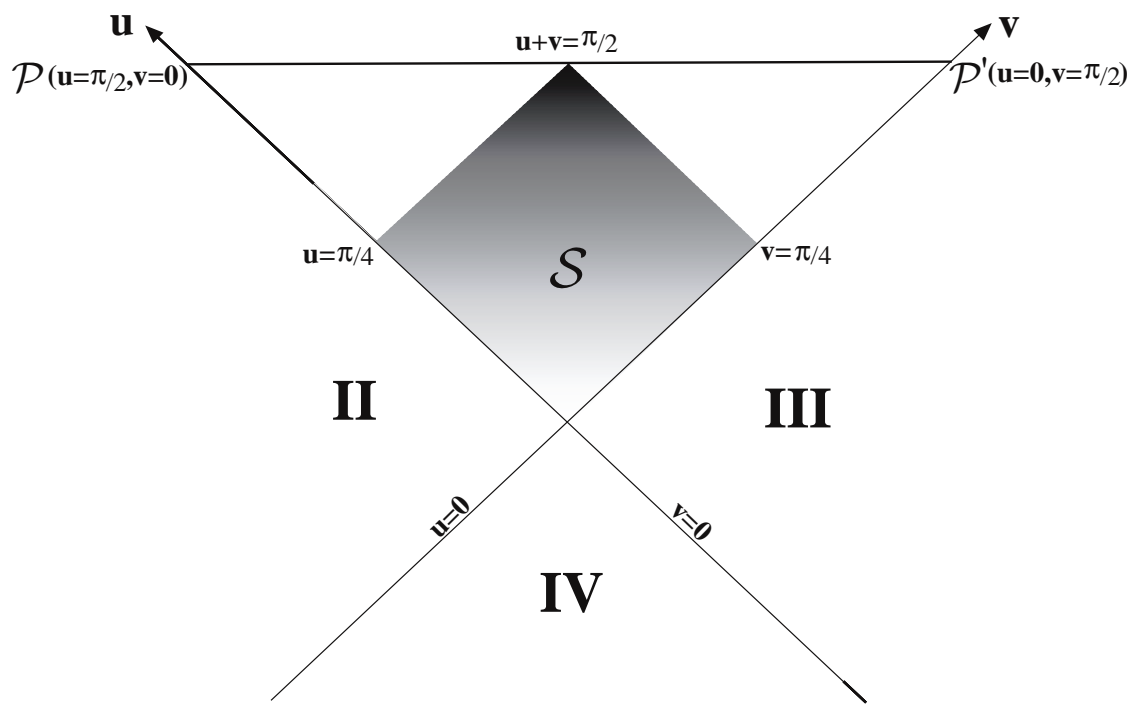
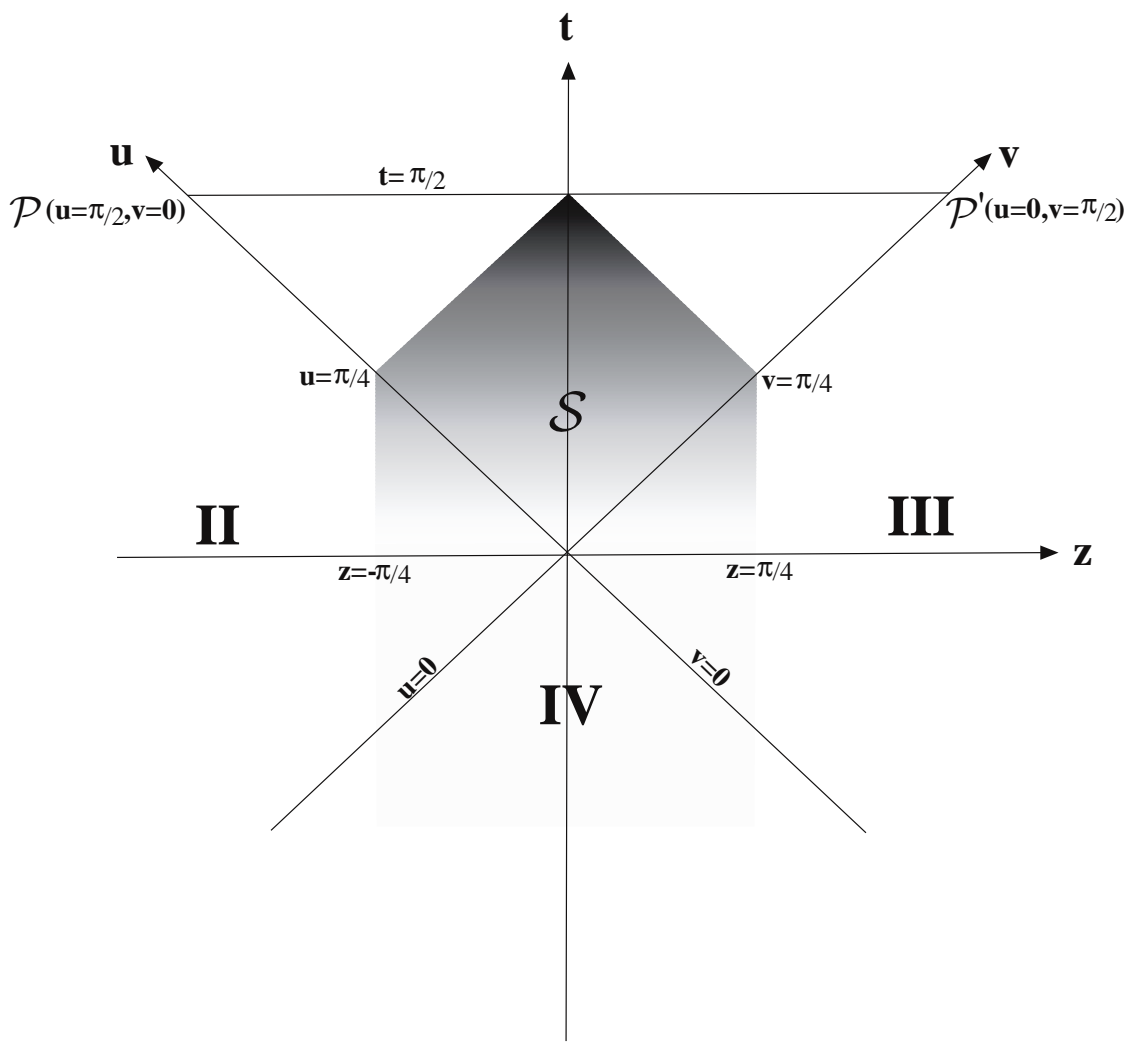
\end{document}